\documentstyle[pra,aps,twocolumn,epsf]{revtex}
\newcommand{\be}{\begin{equation}}
\newcommand{\ee}{\end{equation}}
\newcommand{\bea}{\begin{eqnarray}}
\newcommand{\eea}{\end{eqnarray}}

\newcommand{\dd}{\mbox{d}}
\begin{document}
\draft

\title{Large Spins in External Fields}
\author{V.~A.~Kalatsky$^a$ and V.~L.~Pokrovsky$^{a,b}$}

\address{
$^a$ Department of Physics, Texas A\& M University, 
College Station, Texas 77843-4242
}
\address{
$^b$ Landau Institute for Theoretical Physics, 
Kosygin str.2, Moscow 117940, Russia
}
\date{\today}
\maketitle

\begin{abstract}
Spectra and magnetic properties of large spins $J$, placed into 
a crystal electric field (CEF) of an arbitrary symmetry point group, 
are shown to change drastically when $J$ changes by $1/2$ or $1$. 
At a fixed field symmetry and configuration of its $N$ extrema 
situated at $p$-fold symmetry axis, physical characteristics of the 
spin depend periodically on $J$ with the period equal to $p$. 
The problem of the spectrum and eigenstates of the large spin $J$ is 
equivalent to analogous problem for a scalar charged particle confined 
to a sphere $S^2$ and placed into magnetic field of the monopole 
with the charge $J$. This analogy as well as strong difference between 
close values of $J$ stems from the Berry's phase occurring in the problem. 
For energies close to the extrema of the CEF, the problem can be formulated 
as Harper's equation on the sphere. 
The $2J+1$-dimensional space of states is splitted into smaller multiplets 
of classically degenerated states. These multiplets in turn are splitted 
into submultiplets of states transforming according to specific 
irreducible representations of the symmetry group determined by 
$J$ and $p$. We classify possible configurations and corresponding 
spectra. 
Experimental realizations of large spins in a symmetric 
environment are proposed and physical effects observable in these 
systems are analyzed.

\end{abstract}

\section{Introduction}
\label{section:Introduction}

Conventional wisdom accepts that large spins or orbital momenta 
$J$ (in units of $\hbar$) are almost classical. In particular, if 
$J\gg 1$, their measurable properties do not change substantially 
if $J$ changes by 1/2 or 1. This common belief was undermined by 
Haldane~\cite{Haldane:1} who demonstrated that the ground-state and 
spectrum of the low-energy states in one-dimensional spin chains 
are absolutely different for integer and half-integer spins. 

In this paper we show that similar phenomena can be observed 
on the level of an individual spin placed into external electric field. 
If the field possesses high symmetry (cubic or icosahedral) 
the distinction between spins becomes more subtle. 
For example, in the case of cubic symmetry not only integer spins differ 
from half-integer (this difference is intuitively obvious due to 
the Kramers degeneracy), but the remainder at division of the spin by 
four occurs to determine the spectrum and degeneracy of the low-lying 
states. These striking differences can be found in experiment 
either by spectral analysis or by magnetic measurements. 
We will show that spins 1000, 1001, and 1002 placed into a cubic 
environment have 100\% different magnetic susceptibilities at 
low temperature. Moreover, we will show that a kind of randomness appears 
in properties of large spins in some cases and variation of large spins 
by one can change magnetic and spectral properties in incontrollable way. 

Certainly, the conventional wisdom we started with is presumably correct. 
It is wrong only in a very small range of energy or temperature, the 
smaller the larger is $J$. Nevertheless, as it already happened with 
the Haldane theory, these deviations from classical behavior may be 
important for the experiment. 

The source of all these peculiarities is the Berry's phase. 
Physically, it is associated with the fact that, when the 
classical rotator moves on its unit sphere, it simultaneously rotates 
around its axis. The rotation phase distinguishes the rotator from 
a quantum or classical particle confined on a sphere. The rotator 
problem can be reduced to the particle problem, but the representing 
particle must have an electric charge of unity and must be subjected 
to the homogeneous magnetic field of a monopole with the magnetic 
charge $J$ placed into the center of the sphere. In quantum mechanics 
$J$ accepts integer and half-integer values. 

This paper is composed as follows. In the next section we introduce 
quasi-classical description of large spins. The Berry's phase, 
Berry's connection, and reduction to the problem of a charged particle 
in the monopole field are considered in Section~\ref{section:Berry}. 
In the fourth section we perform the group analysis of the problem. 
The fifth section contains the derivation of the low-energy spectrum 
and magnetic properties of large spins. We separated the case of random 
levels in Section~\ref{section:Random}. Numerical calculations for 
a special potential in a wide range of spin values are given in 
Section~\ref{section:Numerical}. In Section~\ref{section:Experiment} 
we propose experimental realizations of large spins. Our conclusions 
can be found in Section~\ref{section:Conclusion}.

Brief reports on a part of this work were published 
earlier~\cite{tetr,cube}.
\section{Quasi-classical description of large spins}
\label{section:Quasiclassics}

The classical image of a large spin is the classical rotator, 
{\em i.e.}, a vector with a fixed length $J$. 
Its position is determined by 
two spherical coordinates $\theta$ and $\phi$. Sometimes coordinates 
$J_z=J\cos\theta$ and $\phi$ are more convenient since they have 
a simple Poisson brackets: $\{J_z,\phi\}=1$. Classical motion 
is determined by the Hamiltonian:
\be
{\cal H}=f({\bf J})-{\bf h}\cdot{\bf J},
\label{hamilton}
\ee
where ${\bf h}$ is magnetic field (with a precision of a constant factor) and 
$f({\bf J})$ is an arbitrary function of ${\bf J}$, invariant with respect to 
inversion: ${\bf J}\rightarrow-{\bf J}$. The latter requirement is 
equivalent to the time reversal symmetry~\cite{newma71}. Together with 
the standard Poisson brackets $\{J_i,J_j\}=\varepsilon_{ijk}J_k$ the 
Hamiltonian~(\ref{hamilton}) contains full information on classical 
spin dynamics. Periodical trajectories on the sphere can be quantized 
according to the Bohr quantization rule:
\be
\oint J_z(\phi,E)\mbox{d}\phi=(n+\gamma_{\rm B})\pi,
\label{Bohr}
\ee
where $J_z(\phi,E)$ can be found from equation $f({\bf J})=E$ with 
the substitution: $J_x=\sqrt{J^2-J_z^2}\cos\phi$, 
$J_y=\sqrt{J^2-J_z^2}\sin\phi$, and $\gamma_{\rm B}$ is a constant. 

Let us first consider general properties of spin trajectories in zero 
magnetic field. The function $f({\bf J})$, being continuous on the 
sphere, has at least two minima and two maxima. If the external crystal 
field has a non-trivial symmetry group, the number of equivalent 
minima is larger. For example, it can be equal to 4 for tetragonal symmetry, 
6 for hexagonal symmetry. In the case of cubic symmetry it can be 
6, 8, or 12 (directed along 4-, 3-, and 2-fold axes respectively). 
The number of equivalent minima for icosahedral symmetry can be 
12, 20, and 30 (directed along 5-, 3-, and 2-fold axes respectively).
We considered the situations when extrema are located 
in the symmetrical positions. In principal, it is possible that 
they are in more general asymmetric positions. 

Classical trajectories can be separated into two classes: 
``localized'' and ``delocalized''. If energy is close enough to the 
minimum (maximum) of $f({\bf J})$, 
the trajectories are confined in a vicinity 
of one of the minima (maxima). We call such trajectories localized. 
In the intermediate region of the energy trajectories are ``delocalized'', 
they are not confined near any of the extrema. It is obvious that 
delocalized trajectories are highly model-dependent, {\em i.e}, they 
depend on a specific form of $f({\bf J})$. Localized trajectories are 
much more universal: they depend only on the symmetry and on the positions 
of the minima. The same remark is correct with respect to 
quantized levels: low-lying levels, close to $f_{\min}$, or 
almost maximal values of energy, close to $f_{\max}$, have universal 
features, whereas levels in between are rather non-universal. 
Therefore, further we will study only a part of the spectra close 
to $f_{\min}$ or $f_{\max}$. Note that the spectrum of the quantum problem 
is discrete and limited by $f_{\min}$ and $f_{\max}$. 

Before we proceed to detailed study of these levels let us make an 
important remark. For any fixed $J$ and any given $f({\bf J})$ the 
quantum problem consists in the diagonalization of $(2J+1)\times(2J+1)$ 
matrix. Therefore, 
the question arises whether the general theory is necessary. 
The answer is yes. First of all because no reliable information about 
function $f({\bf J})$ is available. We present here general facts, 
independent on specific form of $f({\bf J})$, but only on its symmetry 
group and specific configuration of the extrema. The only requirement for 
our theory is $J\gg1$.

\begin{figure}
  \centerline{
\epsffile{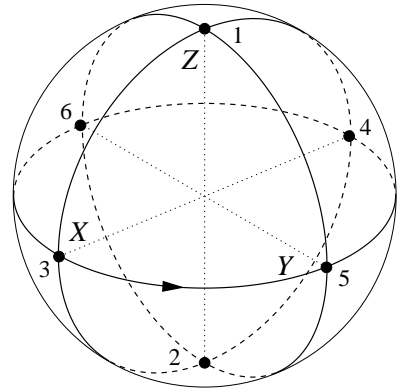}}
  \caption{Tunneling trajectories of the spin (single paths). 
The 6-fold configurations of ${\bf O}$.}
  \label{fig:1}
\end{figure}

Thus, classically a localized 
stationary state is multiply ($N$-fold) degenerate.  
Quantum fluctuations provide a finite radius for each of these states which 
can be enumerated as $|1\rangle,\,|2\rangle,\,\ldots,\,|N\rangle$. 
For considered large $J$ all these states are oscillatory ones within the 
precision $1/J$. More subtle, but not least essential, quantum effect is 
the tunneling between these states. 
The tunneling amplitude between two states $|i\rangle$ and $|j\rangle$, 
$i\not=j$, is exponentially small 
$w_{ij}\propto\exp(-c_{ij}J)$, 
where $c_{ij}$ are constants for a given $f({\bf J})$. 
Therefore, we take into account only tunneling between the 
nearest-neighbor states, {\em i.e.}, the ones with the smallest $c_{ij}=c$, 
and neglect tunneling between more remote states with $c_{ij}>c$. 
To estimate the value of $c$, we need to specify the Hamiltonian. 
For simplicity we consider the case of the cubic symmetry with the 
Hamiltonian:
\be
{\cal H}_1^{\bf O}=-a(J_x^4+J_y^4+J_z^4),
\label{hamilton1}
\ee
where $a>0$ is a constant. 
The minimum value of ${\cal H}_1^{\bf O}$ is 
$E_{\min}=-aJ^4$. There are six minima corresponding to the directions 
of the 4-fold axes: $(\pm J,0,0)$, $(0,\pm J,0)$, $(0,0,\pm J)$. 
Let us consider, for example, tunneling between minima 
$(J,0,0)$, $(0,J,0)$. By symmetry the tunneling trajectory is the smaller 
arc of the big circle passing through these points (Fig.~\ref{fig:1}). 
Putting ${\cal H}_1^{\bf O}=E_{\min}$ we find from eqn.~(\ref{hamilton1}): 
\be
J_z(\phi)=\pm iJ\sqrt{\frac{1-\cos4\phi}{7+\cos4\phi}}.
\ee
The tunneling amplitude is proportional to the exponent;
\be
w\propto\exp(i\int_0^{\pi/2} J_z(\phi)\dd\phi)=\exp(-(J\ln3)/2)=e^{-0.55J}
\label{tunneling}.
\ee

\begin{figure}
  \centerline{
\epsffile{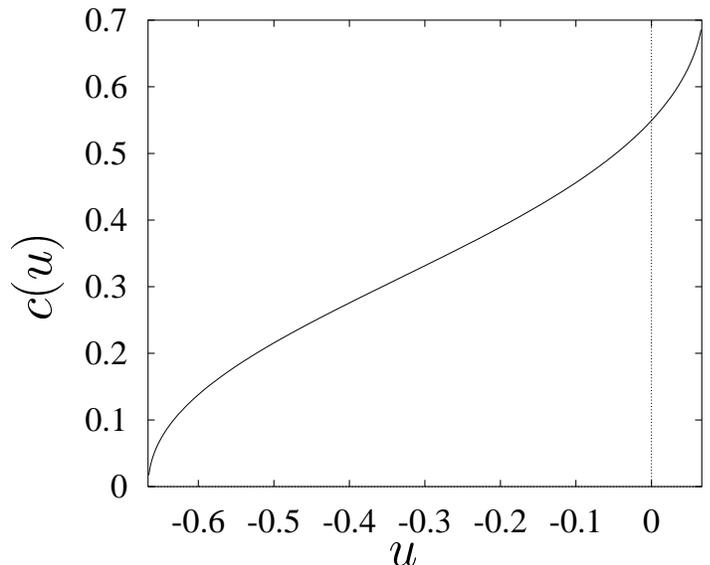}}
  \caption{$c(u)$ for the 6-fold configuration of ${\bf O}$. 
Region of the single tunneling path regime.}
  \label{fig:2}
\end{figure}

For a more realistic Hamiltonian
\be
{\cal H}_2^{\bf O}={\cal H}_1^{\bf O}-
b(J_x^6+J_y^6+J_z^6+30J_x^2J_y^2J_z^2),
\label{hamilton2}
\ee
the exponential factor in the tunneling amplitude 
is $\exp(-c(u)J)$, where $c(u)$ is a function of the ratio 
$u=bJ^2/a$. 
The graph of $c(u)$ is shown in Fig.~\ref{fig:2} 
for values of $u$ in the interval $-2/3< u <1/15$ ($a>0$), where 
the tunneling path passes along the geodesics. 
In the region $1/15< u< 3$ ($a>0$), 
the six minima are still global, however, the tunneling trajectories 
(there are two of them due to the symmetry) deviate from 
the geodesics (see Fig.~\ref{fig:3a})
and the estimation of the exponent becomes more complicated. 
Effects of the multiple path tunneling, for the case of 
the octahedron configurations, will be considered 
in Section~\ref{section:Spectrum}. 
A numerical analysis of the Hamiltonian~(\ref{hamilton2}) 
and comparison to the predictions of the semiclassical  
approximation is given in Section~\ref{section:Numerical}.

\begin{figure}
  \centerline{
\epsffile{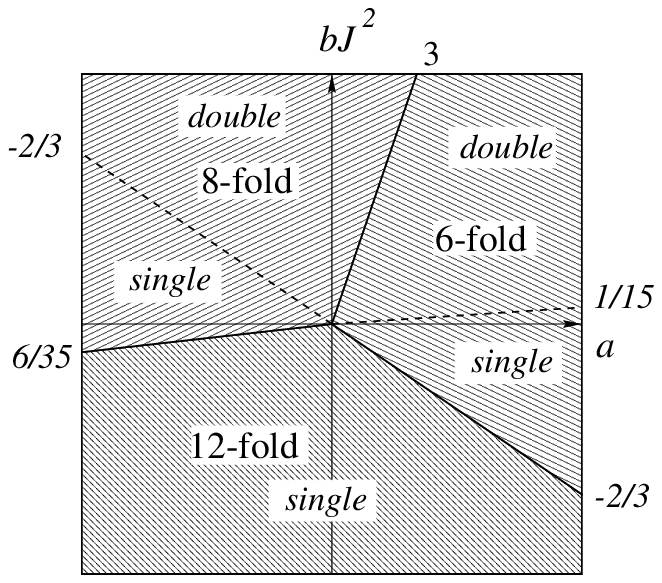}}
  \caption{Phase diagram of Hamiltonian~(\ref{hamilton2}) 
in $bJ^2$-$a$ plane. The dashed lines separate the regions of 
the single and double path tunneling. The numbers on the periphery are 
the slopes of the corresponding lines.}
  \label{fig:3}
\end{figure}

Another important feature of the Hamiltonian~(\ref{hamilton2}) is that, 
depending on signs of $a$ and $b$ and the parameter $u$, it displays 
6, 8, or 12 minima. The phase diagram for this, important for applications, 
Hamiltonian is shown in Fig.~\ref{fig:3}. 
On the boundaries of different ``phases'' different groups of minima 
become equal each other. Then, in the quantum problem, the degeneracy 
of the ground state increases, for example, from 6 to 14. Therefore 
one can expect some singularities in close vicinity of the boundaries. 

For the case of the icosahedral symmetry the simplest Hamiltonian 
is:
\bea
\label{hamilton3}
&&{\cal H}_1^{\bf Y}=-a(J_x^6+J_y^6+J_z^6+30J_x^2J_y^2J_z^2-3\sqrt5\times \\
&&(J_x^2J_y^2(J_x^2-J_y^2)+J_y^2J_z^2(J_y^2-J_z^2)+
J_z^2J_x^2(J_z^2-J_x^2))), \nonumber
\eea
The minimum value of ${\cal H}_1^{\bf Y}$ is $aJ^6/5$ ($a<0$). 
There are 12 minima corresponding to the vertices of an 
icosahedron (directions of the 5-fold axes): 
$J(\alpha,\beta,0)$, $J(0,\alpha,\beta)$, 
$J(\beta,0,\alpha)$, where $\alpha^2=(5+\sqrt5)/10$ and 
$\beta^2=(5-\sqrt5)/10$.
A calculation similar to the one for ${\cal H}_1^{\bf O}$ gives 
the exponential part of the tunneling amplitude $\exp(-0.28J)$.
Positive values of $a$ yield the 20-fold configuration with 
the minima along the 3-fold axes. 

Addition of the next non-trivial invariant of the icosahedron group, 
a polynomial of the 10-{\em th} order over ${\bf J}$, 
allows the configuration with 30 minima along the 2-fold axes.

The tunneling partly lifts the classical degeneracy. 
What was the $N$-fold degenerate state without tunneling is splitted 
into a  multiplet of sub-levels separated by exponentially small energy 
intervals $\propto\exp(-cJ)$, whereas the distances between different 
multiplets are proportional to $1/J$. Each sub-level in the multiplet 
corresponds to a finite-dimensional subspace of states transforming 
according to an irreducible representation of the symmetry group. 
However, as we mentioned already, the realization of this group 
and the spectrum for the rotator is very different from those 
for a quantum particle confined on a sphere. Anyway the problem is reduced 
to diagonalization of a square matrix of the rank $N$ (classical 
degeneracy of the level) with non-zero matrix elements between geometrically 
closest states only. We neglect the tunneling between more remote 
states (non-nearest-neighbors) unless otherwise stated. 

\section{Berry's Phase, Berry's Connection.}
\label{section:Berry}

In the framework of quasi-classical spin dynamics the spin is 
treated as a rigid vector fixed by its direction ${\bf n}$. 
The closest quantum analog is the so-called coherent state 
$|{\bf n}\rangle$ which is defined as an eigenstate of 
operator ${\bf n}\cdot{\bf J}$ with the maximal eigenvalue $J$.
Such a state has minimal uncertainty 
of the spin components transverse to the spin 
quantization axis~\cite{Loss2}. 
An explicit construction for the coherent state reads~\cite{tetr}: 
\be
|{\bf n}\rangle=\exp(iJ_z\phi)\exp(iJ_x\theta)\exp(-iJ_z\phi)|\hat{z}\rangle,
\label{coher-state}
\ee
where $\theta$ and $\phi$ are spherical coordinates of ${\bf n}$;
$|\hat{z}\rangle$ is the coherent state with the direction of quantization axis
along $z$-axis.
This definition assures single-valuedness of the spin wavefunction. 
An adiabatic motion of classical spin ${\bf n}(t)$ 
can be described by the coherent
state $|{\bf n}(t)\rangle$ accompanied with 
a phase factor $e^{i\gamma}$ of purely 
geometrical origin \cite{Berry}. 
Namely, if the spin moves adiabatically along any path $l$
on the unit sphere $S^2$ of ${\bf n}$, the geometrical phase $\gamma (l)$  
is equal to a linear integral:
\be
\gamma=\int_l {\bf A}
\label{g_phase}
\ee
The local change of the phase is 
described by Berry's connection $A_\mu = \langle{\bf n} 
|i\partial/\partial x^\mu |{\bf n}\rangle$. This vector 
field has two components on $S^2$:
\bea
\label{connection}
A_\theta &=& \langle {\bf n} | i\frac{\partial}{\partial\theta}
| {\bf n} \rangle= 0, \\ \nonumber
A_\varphi &=& \frac{1}{\sin\theta}\langle {\bf n} | 
i\frac{\partial}{\partial\varphi}
| {\bf n} \rangle =J\frac{(1-\cos\theta)}{\sin\theta} \ .
\eea
The connection $\bf A$, as well as the geometric phase, is not gauge invariant.
At a local gauge transformation  
$|{\bf n}\rangle\rightarrow\exp(i\lambda({\bf n}))|{\bf n}\rangle$
they are transformed as follows: 
${\bf A}\rightarrow{\bf A}+{\bf d}\lambda$, 
$\gamma\rightarrow\gamma+\lambda_f-\lambda_i\,$, where 
$\lambda$ is an arbitrary differentiable function on $S^2$, 
$i$ and $f$ are its values at 
the initial and final points of path $l$ respectively. 
However, the phase becomes gauge-invariant if the path is 
closed: $l=\partial c$, where $c$ is a surface supported by $l$. In this case:
\be
\gamma(c)=\int_{\partial c}{\bf A}=\int_{c}{\bf dA}=
J\int_{c}\sin(\theta)\dd\theta\dd\phi=J\Omega \ ,
\label{berry_p}
\ee
where $\Omega$ is the solid angle subtended by $\partial c$ at the 
origin of the unit sphere. 
The integrand in~(\ref{berry_p}) is the field-strength 
${\bf B}=J\hat{\bf r}/r^2$. 
This field is identical to magnetic field produced on $S^2$ 
by Dirac's magnetic monopole with the charge $J$ 
located in the center of the sphere.
Thus, following Berry, we formulated the problem of a localized large spin 
in terms of a scalar charged particle confined
on the sphere in the field of magnetic monopole. 

In the presence of a crystal field a further simplification 
becomes possible. As it was shown in 
Section~\ref{section:Quasiclassics}, it leads to the localization 
of the low-energy states near the ``easy'' directions or minima 
of the field and lowers the dimensionality from 
$2J+1$ to $N$, where $N$ is the number of the easy positions.   
The spin trapped near one of the easy 
directions can tunnel to the neighboring minima. 
The tunneling trajectories are solutions of the classical equations
of motion with imaginary time or velocity. The amplitude $w_{ij}$ for
the tunneling from the state $|i\rangle$ to a neighboring state $|j\rangle$
can be written as $w_{ij}=w\exp{i\phi_{ij}}$. Here $w$ is a real, exponentially
small factor (see its calculation in Section~\ref{section:Quasiclassics}) 
and $\phi_{ij}$ is the 
Berry's phase along the tunneling trajectory connecting the points $i$
and $j$. 

The set of Berry's phases $\phi_{ij}$ along the tunneling trajectories 
$\{i,j\}$ connecting extrema labeled by $i$ and $j$ must satisfy a set of
equations. Namely, let us consider a plaquette $c$ on the sphere bounded by
$k$ tunneling paths $\{i_1,i_2\}, \{i_2,i_3\}...\{i_k,i_1\}$. Then:
\be
\sum_m \phi_{i_m,i_{m+1}}=\gamma (c)=J(\Omega(c)\pmod{4\pi})
\label{plaquette}
\ee
where $i_{k+1}=i_1$ and $\Omega (c)$ is the solid angle subtended by the 
contour $\partial c$. The system (\ref{plaquette}) is extended over all
independent plaquettes. Without loss of generality it is possible 
to consider equations~(\ref{plaquette}) only for 
minimal (elementary) plaquettes, {\em i.e.}, 
plaquettes of the minimal non-zero area whose boundaries do not 
have self-intersections. 
Equations (\ref{plaquette}) do not define
the phases $\phi_{ij}$ unambiguously. There remains a freedom of a 
discrete gauge
transformation $\phi_{ij}\rightarrow \phi_{ij}+f_i-f_j$ containing N real
parameters $f_i$. One of them can be treated as a common phase factor and is
inessential. The Schr\"odinger equation in this representation reads:
\be
{\cal H}|\psi\rangle\,=\,E|\psi\rangle
\label{schroedinger}
\ee
where $|\psi\rangle=\sum_{j=1}^Nc_j|j\rangle$ is a vector in the
$N$-dimensional space spanned onto the basis $|j\rangle$,  
$j=1,2...N$ and ${\cal H}$ is an $N\times N$ 
matrix whose diagonal components are equal to a single-well
energy level and nondiagonal elements are $\{{\cal H}\}_{ij}=w_{ij}$. 
Further we put the diagonal matrix elements of $\cal H$ to be zero. 
Then equation
(\ref{schroedinger}) can be rewritten in the vector form:
\be
\sum_{j=1}^Nw_{ij}c_j\,=\,Ec_i;\,\,\,\,w_{ij}=we^{\phi_{ij}}.
\label{vector}
\ee
Eqn. (\ref{vector}) is obviously invariant with respect to the discrete gauge
transformation $w_{ij}\rightarrow w_{ij}e^{i(f_i-f_j)};\,\,c_j\rightarrow
c_je^{if_j}$. Therefore, any set of phases $\phi_{ij}$ satisfying eqns.
(\ref{plaquette}) can be used to find the spectrum and the eigenstates.

We have seen already that the problem of the quasi-classical 
spin is equivalent to the problem of a charged particle 
confined on the sphere $S^2$ in the 
homogeneous magnetic field of the monopole. 
It is a direct spherical analog to the problem of a charged 
particle moving on a plane in a homogeneous magnetic
field, perpendicular to the plane. 
Restricting ourselves with the localized states,
we consider a problem which planar analog is the problem of 
a charged particle living on a 2-$D$ lattice placed into 
homogeneous magnetic field. 
It is known as Harper equation~\cite{Harper}. 
The main difference from this famous problem studied by 
Harper, Azbel, Hofstadter, Thouless, Wiegmann and many other 
authors~\cite{Harper,Azbel}, is that, in our case the lattice 
is embedded into a sphere which
is a compact manifold, in contrast to the planar case. Nevertheless, many
features of the Harper equations will be encountered here, {\it e.g.}, sudden
variations in spectrum at a transition from a rational to an irrational flux
through an elementary plaquette.


The initial Hamiltonian ${\cal H}({\bf J})$ is assumed to possess 
a point group symmetry. It should be noted that ${\cal H}({\bf J})$ 
is invariant with respect to the inversion transformation: 
${\bf J}\rightarrow -{\bf J}$, whereas 
the reduced effective Hamiltonian is not. 
The reason is that this invariance which stems from the time-reversal 
symmetry cannot be extended onto the quantum permutation 
relations: $[J_j,J_k]=\frac{\hbar}{i}\epsilon_{jkl}J_l$. 
The time reversal requires also anti-linear transformation of 
the state-vectors~\cite{Sakurai} which cannot be incorporated into 
linear symmetry group.
Thus all groups of 
transformations under study consist of rotations only.
The point groups in 3-dimensions have been studied thoroughly
(see for example~\cite{LL-QM}). A special interest will be paid to the 
following point groups: $D_n$, $n=2,\,4,\,6$, $O$ (octahedron), 
and $Y$ (icosahedron). 

In the next section we show that the action of the symmetry 
transformations onto the effective Hamiltonian is not trivial 
due to the Berry's phases.


\section{Group theory analysis.}
\label{section:Group}

\subsection{Construction of the main representation.}
\label{subsection:main_representation}

Let $\bar{G}$, a discrete subgroup of $SO(3)$, 
be the point group of the crystal field, 
{\em i.e.}, the point group leaving function $f({\bf J})$ invariant. 
It always includes the space inversion $I$ as a consequence of 
the time-reversal symmetry. Also, we introduce a subgroup $G$ 
of the full symmetry group $\bar{G}=G\times C_i$ ($C_i=\{E,\,I\}$) 
which includes rotation elements only. Further we employ the 
notation ``symmetry group'' namely for $G$. 
Each group $G$ has several sets of equivalent symmetric 
directions defined by the 
intersection of equivalent $p$-fold symmetry axes 
with the unit sphere. Let us denote such a configuration 
${\cal C}(G,p)$ and corresponding number of symmetry directions $N(G,p)$ 
(we denoted it earlier as $N$). 
It can be readily seen that $N(G,p)=|G|/p$, where $|G|$ is 
the rank of the group $G$, {\em i.e.}, the number of its elements. 
The set of $N=N(G,p)$ localized states $|k\rangle$ 
corresponding to the configuration $C(G,p)$ is the vector space for 
a linear unitary representation of the group $G$. 
This representation depends also on $J$. 
Let us call it the main representation and denote it $W(G,p,J)$. 
Its dimensionality is obviously $N(G,p)$. 
For $J=0$, $W(G,p,J)$ is a matrix 
representation of some subgroup $P$ 
of the permutation group $S_N$. 
Each element $g$ of $G$ can be put in one-to-one 
correspondence to a permutation ${\cal P}(g)\in P$. 
The Hamiltonian of the system is invariant under their action. 
For $J\not=0$ or $J$ not equivalent to 0, the problem becomes 
quite peculiar since, due to Berry's phase factors, the Hamiltonian 
is no longer invariant under the action of the transformations ${\cal P}(g)$:
\be
{\cal P}{\cal H}{\cal P}^{T}={\cal H}'\not={\cal H}.
\ee
Hamiltonian ${\cal H}'$ differs from ${\cal H}$ by a gauge transformation. 
Therefore, it is possible to append such a gauge transformation 
${\cal U}\in U$ (unitary diagonal matrix) to each rotation 
that the Hamiltonian remains unchanged:
\be
{\cal U}{\cal P}{\cal H}{\cal P}^{T}{\cal U^{\dagger}}={\cal H}.
\label{invariant_H}
\ee
Thus, a proper representation $W(G,p,J)$ of the symmetry group for 
large spin $J$ or for the Harper's equation on the sphere 
consists of operators  
${\cal W}(g)={\cal U}(g){\cal P}(g)$. 
Since multiplication of each ${\cal W}(g)$ by an arbitrary phase factor 
does not violate Eq.~(\ref{invariant_H}), 
the matrices in $W(G,p,J)$ constitute a projective 
representation of $G$ in general, that is:
\be
{\cal W}(g_1){\cal W}(g_2)=c(g_1,g_2){\cal W}(g_1g_2),\,\,
g_1,\,g_2\in G,
\ee
where $c(g_1,g_2)$ is a function on $G\times G$ 
with values in $U(1)$ (2-dimensional cochain). 

Now, a question arises whether the factor set $c(g_1,g_2)$ is 
equivalent to the trivial one: 
$c'(g_1,g_2)=1$ for any $g_1,\,g_2\in G$, 
as it is for the case $J=0$. By definition, two factor sets 
$c$ and $c^{\prime}$ are equivalent if there exists a function 
$b(g)$ on $G$ 
with values in $U(1)$ (1-dimensional cochain) such that:
\be
c(g_1,g_2)=\frac{b(g_1)b(g_2)}{b(g_1g_2)}c^{\prime}(g_1,g_2). 
\label{equivalence}
\ee

We checked for finite groups $G \subset SO(3)$ that 
equation~(\ref{equivalence}) with $c^{\prime}(g_1,g_2)=\pm 1$ is 
really satisfied. In mathematical language it means,  
that the cochain $c$ is a cocycle but not a 
coboundary for a half-integer spin and it is a coboundary for an 
integer spin~\cite{homology}. 
Therefore, the factor set is non-trivial in general. 
It is equivalent to the multiplicative factors $\{\pm1\}$ 
(isomorphic to ${\bf Z}_2$) which is a consequence of 
the Dirac quantization: $2J=n$, $n\in{\bf N}$. 
This structure of the factor set  might have been anticipated since 
the parameter space of an arbitrary spin is not just 
$SO(3)$ but its universal covering group $SU(2)$ 
which can be obtained as a non-trivial extension of the former one:
$1\rightarrow{\bf Z}_2\rightarrow SU(2)\rightarrow SO(3)\rightarrow 1$. 
In our case, spin in CEF, the proper group of symmetries is $G$ extended 
by ${\bf Z}_2$: 
$1\rightarrow{\bf Z}_2\rightarrow \tilde{G}\rightarrow G\rightarrow 1$.
Instead of dealing with the projective representations of $G$, 
one can work with the linear representations of $\tilde{G}$. 
An explicit construction of $\tilde{G}$ will be given later in this 
section. 
In other language we must consider double-valued representations of 
$G$~\cite{LL-QM} for half-integer $J$.

The representation $W(G,p,J)$ was constructed for a particular gauge, 
however, one can easily find the required representation if 
the Hamiltonian undergoes a gauge transformation: 
$$
{\cal U}{\cal H}{\cal U}^{\dagger}={\cal H}'.
$$
Then a corrected representation leaves the Hamiltonian invariant:
$$
{\cal W}'={\cal U}{\cal W}{\cal U}^{\dagger}. 
$$


\subsection{Classification of configurations.}
\label{subsection:classification}

Generally speaking, the $N$-dimensional main representation 
is reducible.
To perform the reduction of the main representation we need 
to find its characters. 
They are found explicitly in Appendix~\ref{appendix:characters}. 
Here we issue final results.
For $W(G,p,J)$, elements with non-zero characters are: 
identity $E$, the rotation through an angle of $2\pi$ about 
an arbitrary axis $Q$ and rotations $C_p^q$ about the $p$-fold 
axes.
\bea
\chi(E)&=&N;\,\,\chi(Q)=N(-1)^{2J};\,\,
\chi(C_p^q)=2\cos(\frac{2\pi Jq}{p});  \nonumber \\
\chi(C_p^qQ)&=&2(-1)^{2J}\cos(\frac{2\pi Jq}{p});\,\, (q=1,\ldots, p-1)
\eea

Now we proceed to consideration of different point groups and their
configurations of extrema. 

\subsubsection{Configurations of the octahedron group ${\bf O}$}
\label{O}

Here, we classify possible configurations ${\cal C}({\bf O},p)$ 
of the octahedron symmetry group ${\bf O}$. 
In general, {\em i.e.}, without 
accidental degeneracy, the minima (maxima) of the 
potential are located either on the equivalent 
symmetry axes of the cube or completely away from them (asymmetrically): 
\begin{description}
\item[${\cal C}({\bf O},4)$] 
three axes of the fourth order passing through the centers of 
opposite faces, $N=6$.
\item[${\cal C}({\bf O},3)$] 
four axes of the third order passing through opposite corners, $N=8$.
\item[${\cal C}({\bf O},2)$] 
six axes of the second order through the midpoints of opposite 
edges, $N=12$.
\item[${\cal C}({\bf O},1)$] 
none of the symmetry axes passes through the minima $N=24$ or $N=48$.
\end{description}
The representations $W({\bf O},p,J)$ of the octahedron group 
acting on the spaces of states corresponding to the above described 
configurations are respectively 6-, 8-, 12-, and 24(48)-dimensional. 

For a configuration ${\cal C}({\bf O},p)$ only elements $C_p^q$ 
have non-zero characters which were calculated earlier. 
They must be divided into classes of conjugate elements. 
The classes with non-zero characters (except $E$ and $Q$) are:
six rotations $C_4$ and $C_4^3$, 
and three rotations $C_4^2$ for ${\cal C}({\bf O},4)$; 
eight rotations $C_3$ and $C_3^2$ for ${\cal C}({\bf O},3)$; 
six rotations $C_2$ for ${\cal C}({\bf O},2)$; 
and none for ${\cal C}({\bf O},1)$.

The characters are periodic functions of $J$ with the period equal to $p$. 
It means that the multiplicities of the eigenvalues of the 
Hamiltonian have the same periodicity.
The characters are invariant under the transformation 
$J\rightarrow -J \pmod{p}$ (reflection). 

The irreducible components contained in representations 
$W_N$, $N=6,\,8,\,12,\,24$ of the octahedron group are given in 
Table~\ref{tab:2} for values of $J$ inequivalent under 
the translations over $p$ and the reflection.
For simplification we denoted $W({\bf O},p,J)$ as $W_N$, 
where $N=N({\bf O},p)$. 
The irreducible components of $W_{48}$ are not listed since 
there are twice as many of them as those for $W_{24}$. 
This relationship is correct for representation 
$W_{|\bar{G}|}$ of any group $G$. 
The characters of the accidental configurations, such as 
the 14-fold configurations on the boundary between 
${\cal C}({\bf O},4)$ and ${\cal C}({\bf O},3)$, 
are merely sums of the characters of the constituting components 
and can be found from the given tables for the basic configurations. 


\begin{center}
\begin{table}
\caption{Irreducible components of the cubic representations 
$W_N$, $N=6,\,8,\,12,\,24$.}
\label{tab:2}
\begin{tabular}{ccccc}
$J$ &  ${\cal C}({\bf O},4)$ & ${\cal C}({\bf O},3)$ & 
${\cal C}({\bf O},2)$ & ${\cal C}({\bf O},1)$ \\
\hline
0 & $A_1$, $E$, $F_1$ & 
\begin{tabular}{c}
$A_1$, $A_2$ \\ 
$F_1$, $F_2$ \\
\end{tabular} &
\begin{tabular}{c} 
$A_1$, $E$ \\ 
$F_1$, $F_2(2)$ \\
\end{tabular} & 
\begin{tabular}{c}
$A_1$, $A_2$, $E(2)$ \\ 
$F_1(3)$, $F_2(3)$ \\
\end{tabular} \\
1 & $F_1$, $F_2$ & $E$, $F_1$, $F_2$ & 
\begin{tabular}{c}
$A_2$, $E$ \\ 
$F_1(2)$, $F_2$ \\ 
\end{tabular} & \\
2 & $A_2$, $E$, $F_2$ &  &  & \\
\hline
1/2 & $E_1'$, $G'$ & $E_1'$, $E_2'$, $G'$ & $E_1'$, $E_2'$, $G'(2)$ & 
\begin{tabular}{c}
$E_1'(2)$, $E_2'(2)$ \\
 $G'(4)$ \\
\end{tabular} \\
3/2 & $E_2'$, $G'$ & $G'(2)$ &  \\
\end{tabular}
\end{table}
\end{center}
 
\subsubsection{Configurations of the icosahedron group ${\bf Y}$}
\label{Y}

The classification of configurations ${\cal C}({\bf Y},p)$ for 
the icosahedron group of symmetries is  
similar to that of the octahedron group. 
Extrema can be located either along the directions 
of the symmetry axes or asymmetrically:
\begin{description}
\item[${\cal C}({\bf Y},5)$] 
six axes of the fifth order passing through 
opposite corners of the icosahedron, $N=12$.
\item[${\cal C}({\bf Y},3)$] 
ten axes of the third order passing through the centers of 
opposite faces $N=20$. 
\item[${\cal C}({\bf Y},2)$] 
fifteen axes of the second order through the midpoints of opposite 
edges $N=30$.
\item[${\cal C}({\bf Y},1)$] 
none of the symmetry axes passes through the minima $N=60$ or $120$.
\end{description}
The main representations of the icosahedron group acting on the spaces 
of the configurations are respectively 12-, 20-, 30-, and 
60(120)-dimensional. 
The classes with non-zero characters, besides $E$ and $Q$, are:
twelve rotations $C_5^{1,4}$ and twelve rotations $C_5^{2,3}$ 
for ${\cal C}({\bf Y},5)$, 
twenty rotations $C_3^{1,2}$ for ${\cal C}({\bf Y},3)$,
fifteen rotations $C_2$ for ${\cal C}({\bf Y},2)$, 
and none for ${\cal C}({\bf Y},1)$.
The multiplicities of the eigenvalues of the Hamiltonian, 
for a configuration ${\cal C}({\bf Y},p)$, 
have the period $p$. 
The irreducible components contained in representations 
$W_N$, $N=12,\,20,\,30,\,60$ of the icosahedron group are given in 
Table~\ref{tab:4}


\begin{center}
\begin{table}
\caption{Irreducible components of the icosahedron representations 
$W_N$, $N=12,\,20,\,30,\,60$.}
\label{tab:4}
\begin{tabular}{ccccc}
$J$ & ${\cal C}({\bf Y},5)$ & ${\cal C}({\bf Y},3)$ & 
${\cal C}({\bf Y},2)$ & ${\cal C}({\bf Y},1)$ \\
\hline
0 & 
\begin{tabular}{c}
$A$, $F_1$ \\ 
$F_2$, $H$ \\ 
\end{tabular} & 
\begin{tabular}{c}
$A$, $F_1$, $F_2$ \\ 
$G(2)$, $H$ \\ 
\end{tabular} & 
\begin{tabular}{c}
$A$, $F_1$, $F_2$ \\ 
$G(2)$, $H(3)$ \\
\end{tabular} & 
\begin{tabular}{c}
$A$, $F_1(3)$, $F_2(3)$ \\ 
$G(4)$, $H(5)$ \\
\end{tabular} \\
1 & $F_1$, $G$, $H$ & 
\begin{tabular}{c}
$F_1$, $F_2$ \\ 
$G$, $H(2)$ \\ 
\end{tabular} &
\begin{tabular}{c}
$F_1(2)$, $F_2(2)$ \\
$G(2)$, $H(2)$ \\
\end{tabular} & \\
2 & $F_2$, $G$, $H$ &  &  & \\
\hline
1/2 & $E_1'$, $G'$, $I'$ & 
\begin{tabular}{c}
$E_1'$, $E_2'$ \\ 
$G'$, $I'(2)$ \\ 
\end{tabular} & 
\begin{tabular}{c}
$E_1'$, $E_2'$ \\ 
$G'(2)$, $I'(3)$ \\
\end{tabular} & 
\begin{tabular}{c}
$E_1'(2)$, $E_2'(2)$ \\ 
$G'(4)$, $I'(6)$ \\
\end{tabular} \\
3/2 & $E_2'$, $G'$, $I'$ & $G'(2)$, $I'(2)$ &  & \\
5/2 & $I'(2)$ & & & \\
\end{tabular}
\end{table}
\end{center}

\subsubsection{Configurations of ${\bf D}_2$}
\label{D2}

Next, we consider the configurations of three groups of symmetries 
${\bf D}_N$ ($N=2,\,4,\,6$). 
Despite their simplicity, Berry's phase introduces here some interesting 
effects as well. 

The configurations of ${\bf D_2}$ are quite simple:
\begin{description}
\item[${\cal C}({\bf D}_2,2)$] 
one axis of the second order, $N=2$.  
\item[${\cal C}({\bf D}_2,1)$] 
none of the symmetry axes passes through the extrema, $N=4$ or $8$.
\end{description}
The characters of the ${\bf D}_2$ representations and 
the irreducible components contained in representations 
$W_N$, $N=2,\,4$  are given in Table~\ref{tab:5_D2}.

\begin{center}
\begin{table}
\caption{Irreducible components of the ${\bf D}_2$ representations.}
\label{tab:5_D2}
\begin{tabular}{ccc}
$J$ & ${\cal C}({\bf D}_2,2)$ & ${\cal C}({\bf D}_2,1)$ \\
\hline
0 & $A$, $B_3$ & $A$, $B_1$, $B_2$, $B_3$ \\
1 & $B_1$, $B_2$ &   \\
\hline
1/2 & $E'$ & $E'(2)$ \\
\end{tabular}
\end{table}
\end{center}

\subsubsection{Configurations of the tetragonal group ${\bf D}_4$}
\label{D4}

The configurations of ${\bf D_4}$ 
are listed below:
\begin{description}
\item[${\cal C}({\bf D}_4,4)$] 
one axis of the fourth order, $N=2$. 
\item[${\cal C}({\bf D_4},2)$] 
two axes of the second order, $N=4$. 
\item[${\cal C}({\bf D_4},1)$] 
none of the symmetry axes passes through the extrema, $N=8$ or $16$.
\end{description}
The irreducible components contained in representations 
$W_N$, $N=2,\,4,\,8$  of ${\bf D}_4$ are given in Table~\ref{tab:5}.

An interesting conclusion can be drawn from the data in Table~\ref{tab:5}. 
The 2-fold classical degeneracy of the configurations of 
${\cal C}({\bf D}_4,4)$ is not lifted for all but even values of $J$. 
Thus, the tunneling is allowed only for even spins. This result 
cannot be accounted for by Kramers degeneracy, as it was possible 
in~\cite{Loss} for ${\bf D}_2$ configuration, 
and is totally due to the symmetry combined with the Berry's phase. 
Also, it shows importance of the details of the background, {\em i.e.}, 
${\bf D}_2$ (considered in the previous section and in~\cite{Loss}), 
${\bf D}_4$, and ${\bf D}_6$ (considered in the next section) 
groups of symmetries 
have the easy axis (2-fold) configuration, however the tunneling 
is allowed in the ${\bf D}_N$ environment only for $J=0 \!\pmod{N/2}$ 
and is defined by the anisotropy in the plane normal to the easy axis.

\begin{center}
\begin{table}
\caption{Irreducible components of the ${\bf D}_4$ representations.}
\label{tab:5}
\begin{tabular}{cccc}
$J$ & ${\cal C}({\bf D}_4,4)$ & ${\cal C}({\bf D}_4,2)$ & 
${\cal C}({\bf D}_4,1)$  \\
\hline
0 & $A_1$, $A_2$ & $A_1$, $B_1$, $E$ & $A_1$, $A_2$, $B_1$, $B_2$, $E(2)$ \\
1 & $E$ & $A_2$, $B_2$, $E$ & \\
2 & $B_1$, $B_2$ &  & \\
\hline
1/2 & $E_1'$ & $E_1'$, $E_2'$ & $E_1'(2)$, $E_2'(2)$ \\
3/2 & $E_2'$ &  & \\
\end{tabular}
\end{table}
\end{center}

\subsubsection{Configurations of the hexagonal group ${\bf D}_6$}
\label{D6}

Due to similarity of this group with ${\bf D}_4$, we just present 
the data on the ${\bf D}_6$ representations. 

\begin{description}
\item[${\cal C}({\bf D}_6,6)$] 
one axis of the sixth order, $N=2$. 
\item[${\cal C}({\bf D}_6,2)$] 
three axes of the second order, $N=6$. 
\item[${\cal C}({\bf D}_6,1)$] 
none of the symmetry axes passes through the minima, $N=12$ or $24$. 
\end{description}

\begin{center}
\begin{table}
\caption{Irreducible components of the ${\bf D}_6$ representations.}
\label{tab:6}
\begin{tabular}{cccc}
$J$ & ${\cal C}({\bf D}_6,6)$ & ${\cal C}({\bf D}_6,2)$ & 
${\cal C}({\bf D}_6,1)$  \\
\hline
0 & $A_1$, $A_2$ & $A_1$, $B_1$, $E_1$, $E_2$ & 
$A_1$, $A_2$, $B_1$, $B_2$, $E_1(2)$, $E_2(2)$ \\
1 & $E_1$ & $A_2$, $B_2$, $E_1$, $E_2$ & \\
2 & $E_2$ &  & \\
3 & $B_1$, $B_2$ &  & \\
\hline
1/2 & $E_1'$ & $E_1'$, $E_2'$, $E_3'$ & $E_1'(2)$, $E_2'(2)$ , $E_3'(2)$ \\
3/2 & $E_3'$ &  & \\
5/2 & $E_2'$ &  & \\
\end{tabular}
\end{table}
\end{center}
\section{Spectrum}
\label{section:Spectrum}

The group-theoretical analysis of the last section gives 
the number of splitted sublevels in the initial $N$-fold 
multiplet and their degeneracies. In this section 
we find the order of the sublevels and distances between them. 
It requires explicit diagonalization of the reduced Hamiltonian. 
As we show below, the spectrum is much more subtle matter 
than the number and degeneracy of the sublevels. It may 
depend on details of the Hamiltonian. 

We assume that all tunneling paths between 
nearest minima are equivalent, that is, all non-zero tunneling 
amplitudes have equal absolute values $|w|$. 
Consequently, $w$ enters the Hamiltonian as a common multiplier 
and all eigenvalues are multiples of $w$ in zero magnetic field.
The solid angle covered by the minimal non-trivial closed 
path will be assumed known. It is, actually, a constant for all 
configurations but ${\cal C}(G,2)$, $G={\bf O},\,{\bf Y}$, 
where it is a function of some dimensionless combinations 
of the CEF parameters, {\em e.g.}, ratio $u$ in 
Section~\ref{section:Quasiclassics}. 

In some cases, not only in simple ones, such as ${\cal C}({\bf D}_N,N)$, 
there are two tunneling trajectories connecting nearest 
minima. {\em E.g.}, in a vicinity of the boundary between 
6- and 8-fold configurations of the cubic group (see Fig.~\ref{fig:3}), 
the tunneling trajectory deviates from the geodesics connecting 
the minima and, due to the symmetry, there are two trajectories located, 
symmetrically with respect to the geodesics. 
However, the two trajectories can be considered as one 
effective path with the tunneling amplitude of $2w\cos(J\Omega/2)$ 
(see Eq.~(\ref{double-trajectory})), 
where $w$ is the tunneling amplitude of a single path and 
$\Omega$ is the solid angle subtended by the two trajectories. 

Before proceeding to a detailed analysis of the spectra, 
we obtain some relations between eigenvalues of 
the same configuration, but for different $J$. These relations 
are of purely geometric origin~\cite{cube}. 
Let us assume that the parameter space $S^2$ can be covered 
completely and without overlap by 
$s$ congruent plaquettes whose boundaries are the tunneling 
trajectories. 
{\em E.g.}, these are 2 hemispheres for ${\cal C}({\bf D}_N,2)$, 
$N=2,\,4,\,6$, configurations; 
$N$ orange-like segments for ${\cal C}({\bf D}_N,N)$, Fig.~\ref{fig:4}; 
8 curved right-angled triangles for ${\cal C}({\bf O},4)$, Fig.~\ref{fig:1}. 
Each plaquette subtends a solid angle of $4\pi/s$, and Berry's phase for 
each loop is $4\pi J/s$. Then, from (\ref{plaquette}), it follows that 
the spectrum is a periodic function of $J$ with the period 
$s/2$~\footnote{This statement is conventional: it is periodic 
if $w$ does not depend on $J$. However, the ratios of the 
interlevel distances are periodic functions of $J$.}. 
The spectra of systems differing by transformation 
$\gamma(c)\rightarrow-\gamma(c)$ must be identical due to the 
time-reversal symmetry. Hence, all $J$'s are divided into $s/2+1$ 
equivalence classes defined by a set of numbers 
$0,\,1/2,\,1,\ldots,s/4$. A fixed $J$ belongs to the class of equivalence 
labeled by: 
\be 
\min_{n\in{\bf Z}}|J+ns/2|. 
\label{equivalence-r}
\ee
Hereinafter, we will work only with the minimal non-equivalent $J$'s.

In a more general setting, {\em i.e.}, in the presence 
of $n$ different elementary plaquettes, periodicity 
of the spectra depends on the rationality of the flux 
quanta passing through each plaquette: 
if a flux per each plaquette is 
$\Phi_i=J\Omega_i=2\pi JP_i/Q_i$, $i=1,\ldots,n$,  
where $P_i$ and $Q_i$ are mutually prime integers, 
then the period of the spectra is the least common multiple 
of $Q_i$, $i=1,\ldots,n$. Otherwise the spectra are not periodic and 
each $J$ represents a class. 
Thus, if $n=1$ the spectra is always periodic and 
if $n>1$ it is not in general 
(unless an additional symmetry is present). 

An extra symmetry of the spectra can be extracted by considering 
an operation of the change of sign: $w\rightarrow -w$. 
This transformation inverts energy levels inside of each class. 
On the other hand, the spectra 
depend only on gauge invariants 
$w^k\cos(J\Omega_k)$, where $\Omega_k=4m\pi/s$ ($m$ is an integer) 
is the solid angle subtended 
by a closed contour containing $k$ tunneling paths 
and is a multiple of the solid angle 
subtended by the elementary plaquette $4\pi/s$. 
If all closed contours contain even number of the paths ($k$ is even), 
{\em e.g.}, ${\cal C}({\bf O},3)$, ${\cal C}({\bf D}_4,2)$, 
the levels are symmetric inside of each class, that is, 
they come in pairs of opposite sign $\pm E$. 
For example, for eigenvalues of 
${\cal C}({\bf O},3)$ the following relations are satisfied:
$E(J=0;A_1)=-E(J=0;A_2)$, $E(J=0,1;F_1)=-E(J=0,1;F_2)$, 
$E(J=1;E)=0$, $E(J=1/2;E_1')=-E(J=1/2;E_2')$, $E(J=1/2;G')=0$.
If some of the closed contours consist of odd number of the paths, 
{\em e.g.}, ${\cal C}({\bf O},4)$, ${\cal C}({\bf Y},5)$, then 
the simultaneous change of sign $w\rightarrow-w$ and shift 
$J\rightarrow J+s/4$ leaves the invariant combinations unchanged. 
Therefore, each level $E$ in the class of $J$ has its counterpart 
$-E$ in the class of $J+s/4$. 
For example, in ${\cal C}({\bf O},4)$: 
$E(J=0;A_1)=-E(J=2;A_2)$, $E(J=0;F_1)=-E(J=2;F_2)$, 
$E(J=0;E)=-E(J=2;E)$, $E(J=1/2;E_1')=-E(J=3/2;E_2')$, 
$E(J=1/2;G')=-E(J=3/2;G')$.
If $J$ and $J+s/4$ belong to the same equivalence class, their 
spectrum is symmetric, {\em e.g.}, in ${\cal C}({\bf O},4)$: 
$E(J=1;F_1)=-E(J=1;F_2)$.

Further in this section we calculate the spectra for different 
groups of symmetry and configurations. 

\subsection{Spectra of the ${\bf D}_{n}$ ($n=2,\,4,\,6$).}
\label{spectrum-D}

\begin{figure}
  \centerline{
\epsffile{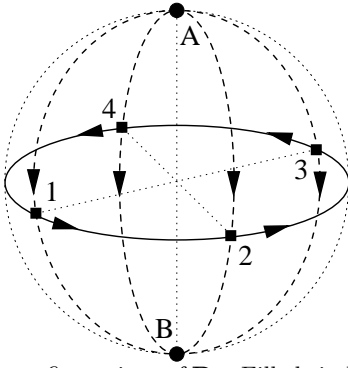}}
  \caption{The configurations of ${\bf D}_4$. 
Filled circles and dashed lines belong to ${\cal C}({\bf D}_4,4)$, and 
filled squares and solid lines belong to ${\cal C}({\bf D}_4,2)$ 
respectively.}
  \label{fig:4}
\end{figure}

The configurations of ${\bf D}_4$ are shown on Fig.~\ref{fig:4}. 
In the case of the ${\bf D}_6$ configurations, there are six minima on 
the equatorial circle (${\cal C}({\bf D}_6,2)$) and six tunneling paths 
connecting the antipodal points (${\cal C}({\bf D}_6,6)$). 
For ${\cal C}({\bf D}_2,2)$, it is just two minima connected by 
two tunneling trajectories. 
The total tunneling amplitude for ${\cal C}({\bf D}_N,N)$, from one 
pole to the other is: 
\be
w\sum_{k=0}^{N-1}\exp(i4\pi kJ/N),
\ee
where we prescribed a phase factor of unity to one of the tunneling paths.
The Hamiltonian is a $2\times2$ matrix with the following eigenvalues: 
\be
E=\left\{
\begin{array}{ll}
\pm&2w\cos(\pi J) \mbox{ for } N=2,  \\ 
\pm&4w\cos(\pi J)\cos(\pi J/2) \mbox{ for } N=4, \\ 
\pm&2w\cos(\pi J)(1+2\cos(2\pi J/3)) \mbox{ for } N=6. 
\end{array}
\right.
\label{D-2-fold}
\ee
In full agreement with the predictions of Section~\ref{section:Group}, 
the paths interfere destructively for all spin values but $J=0\!\pmod{N/2}$.

The case of minimal symmetry ${\cal C}({\bf D}_2,2)$ has been considered 
by D.~Loss {\it et al}~\cite{Loss} earlier. They argued that in the case 
of half-integer $J$ the tunneling amplitudes along the two paths cancel 
each other. One can see from Eqns.~(\ref{D-2-fold}) that, when the number 
of equivalent tunneling paths increases due to the symmetry, such 
a cancelation takes place for integer $J$ as well (with exception 
of $J=0\!\pmod{N/2}$), where the classical degeneracy of the 
ground-state level is 2-fold for all ${\cal C}({\bf D}_N,N)$. 

In the presence of magnetic field the eigenvalues are 
$E(h)=\pm\sqrt{E^2(0)+(hJ)^2}$, where $h=g\mu_{\rm B}H$ and $H$ is 
the component of magnetic field along the easy direction.

For the ${\cal C}({\bf D}_N,2)$ $N=4,\,6$ configurations, the Hamiltonian 
is that of the one-dimensional $N$-site tight-binding model~\cite{tetr}, 
with eigenvalues~\footnote{The label $k$ in Eqn.~\ref{tight-binding} 
does not correspond to the algebraic value of the level.}:
\be
E_k=2w\cos(2\pi(k+J)/N),\,k=0,\,1,\,\ldots,\,N-1. 
\label{tight-binding}
\ee

The magnetic field enters the Hamiltonian as a site-diagonal 
matrix:
\be
{\cal H}_h=-hJ\cos(\phi_h-2\pi l/N),\,l=0,\,1,\ldots,N-1,
\label{Zeeman}
\ee
where $h=g\mu_{\rm B}H$, $H$ is the in-plane component of magnetic field, and 
$\phi_h$ is the angle of this component with respect to the 
easy direction of the CEF labeled by $l=0$. 
The eigenvalues of ${\cal H}+{\cal H}_h$ can be found analytically. 
For ${\cal C}({\bf D}_4,2)$, one finds: 
$$
E^2=2w^2+{\bar h^2\over 2}\pm
\sqrt{4w^4\cos^2(\pi J)+2\bar h^2w^2+{\bar h^4\over 4}\cos^2(2\phi_h)},
$$
where we used $\bar h$ as a shorthand for $hJ$. 
The spectra of ${\cal C}({\bf D}_4,2)$ 
(previously calculated in~\cite{tetr}) and ${\cal C}({\bf D}_6,2)$ 
in magnetic field are given in the limit of small magnetic fields in 
Tables~\ref{tab:7} and~\ref{tab:8} respectively. 
The last column of Tables~\ref{tab:7},~\ref{tab:8} is the low temperature 
magnetic susceptibility which is a readily observable physical quantity. 
The susceptibility saturates to a constant for the 
classes without a magnetic moment in the ground-state (integer spins) 
and has a Curie-like behavior for the ones with a magnetic moment 
in the ground-state (half-integer spins); see Appendix~\ref{appendix:B} 
for details.

In the case of ${\cal C}({\bf D}_N,2)$ configurations, there is 
a spectral difference between integer and half-integer spins only 
which can be ascribed to Kramer's degeneracy. 
In the next section, we consider non-Abelian cases, 
where more complex division on equivalence classes occurs.

\begin{center}
\begin{table}
\caption{Spectrum of ${\cal C}({\bf D}_4,2)$ in magnetic field, 
the limit of small magnetic field, 
and low temperature magnetic susceptibility ($\beta=1/(k_{\rm B}T)$.}
\label{tab:7}
\begin{tabular}{ccc}
J & Eigenvalues & Susceptibility \\
\hline
0   & $\pm(2w+\frac14h^2J^2/w)$, $\pm\frac14h^2J^2\sin(2\phi_h)/w$ & 
$\frac12(gJ\mu_{\rm B})^2/w$ \\
$1/2$ & $\pm(\sqrt2w\pm\frac12hJ+\frac{\sqrt2}{16}h^2J^2/w)$ & 
$\frac14(gJ\mu_{\rm B})^2\beta$ \\
\end{tabular}
\end{table}
\end{center}

\begin{center}
\begin{table}
\caption{Spectrum of ${\cal C}({\bf D}_6,2)$ in magnetic field, 
the limit of small magnetic field, 
and low temperature magnetic susceptibility.}
\label{tab:8}
\begin{tabular}{ccc}
J & Eigenvalues & Susceptibility \\
\hline
0   & 
\begin{tabular}{c}
$\pm(2w+\frac12h^2J^2/w)$ \\ 
$\pm(w-\frac38h^2J^2/w)$ \\ 
$\pm(w+\frac18h^2J^2/w)$ \\
\end{tabular}
& 
$(gJ\mu_{\rm B})^2/w$ \\
$1/2$ & $0\,(2)$, $\pm(\sqrt3w\pm\frac12hJ+\frac{\sqrt3}{12}h^2J^2/w)$ & 
$\frac14(gJ\mu_{\rm B})^2/(k_{\rm B}T)$ \\
\end{tabular}
\end{table}
\end{center}

\subsection{Spectra of the ${\bf O}$ configurations}
\label{spectrum-O}

The cubic symmetries are quite common in nature. 
We will perform a detailed study of the configurations 
of the octahedron group. 
The Hamiltonian for configuration ${\cal C}({\bf O},4)$ 
has the following matrix elements:
\bea
h_{ii}&=&0,\,\,i=1,\,2,\ldots 6, \nonumber \\  \label{Ham_O_6}
h_{ij}&=&0,\,\,|i-j|=1,\,i+j=3,\,7,\,11, \\
|h_{ij}|&=&|w|,\,\,\mbox{for other } 1\leq i,j\leq6\nonumber
\eea
where we adopted the enumeration shown on Fig.~\ref{fig:1}. 
The tunneling trajectories divide the sphere into eight plaquettes. 
Relation~(\ref{plaquette}), written for each plaquette, gives 
eight equations for the phases $\phi_{ij}$, where $\Omega(c)=\pi/2$ 
(an example of a set of the phases for this configuration 
as well as a calculation of the spectra 
is given in Appendix~\ref{appendix:A1}). 
Only seven equations are independent. Given definite phases, 
the diagonalization is straightforward. The eigenvalues can 
be expressed in the following closed form~\cite{cube}:
\bea
\label{exact6}
E_k(J)&=&(-1)^k2w\chi(\pi(J+2k)),\,\,k=0,\ldots,5, \\
\chi(x)&=&\cos\frac{2x}{3}\cos\frac{x}{2}-
\left(\cos^2\frac{x}{3}+
\sin^2\frac{2x}{3}\sin^2\frac{x}{2}\right)^{1/2}. \nonumber
\eea
The ordered spectra of ${\cal C}({\bf O},4)$ are given 
in Table~\ref{tab:9} ($w>0$) 
for the minimal set of $J$'s; the spectra for other $J$'s can be obtained 
by the use of the equivalence relation~(\ref{equivalence-r}). Note 
that the spectra should be inverted if $w$ is negative. 

\begin{table}
\caption{The spectra and the low temperature magnetic susceptibilities 
of ${\cal C}({\bf O},4)$ ($\beta=1/(k_{\rm B}T)$, 
a common factor of $(gJ\mu_{\rm B})^2$ is omitted ).}
\begin{tabular}{lll}
$J$ & Eigenvalues(degeneracies) &Susceptibility\\ \hline
0 & $-2w\,(2)$, $0\,(3)$, $4w\,(1)$  & $1/(3w)$\\
1 & $-2w\,(3)$, $2w\,(3)$            & $\beta/6$ \\
2 & $-4w\,(1)$, $0\,(3)$, $2w\,(2)$  & $1/(6w)$  \\
1/2 & $-\sqrt2w\,(4)$, $2\sqrt2w\,(2)$ & $2\beta/9$\\
3/2 & $-2\sqrt2w\,(2)$, $\sqrt2w\,(4)$ & $\beta/9$ \\
\end{tabular}
\label{tab:9}
\end{table}

The physical difference among the classes is 
manifested when magnetic field is applied.  
The magnetic part of the Hamiltonian in this case is:
$$
{\cal H}_h=
J\,diag(-h_z,\,h_z,-h_x,\,h_x,-h_y,\,h_y).
$$ 
The full Hamiltonian can be easily diagonalized for 
some symmetric direction of the field, {\em e.g.}, along 
easy direction $(1,0,0)$. The direction of the field 
does not influence the low temperature susceptibility 
since the latter is isotropic in a cubic CEF. 
However, individual levels of the ground-state multiplet may have 
anisotropic magnetic susceptibility as well as anisotropic 
magnetization. 
The low temperature magnetic susceptibilities of ${\cal C}({\bf O},4)$ 
are collected in the last column of Table~\ref{tab:9}. 

\begin{figure}
  \centerline{
\epsffile{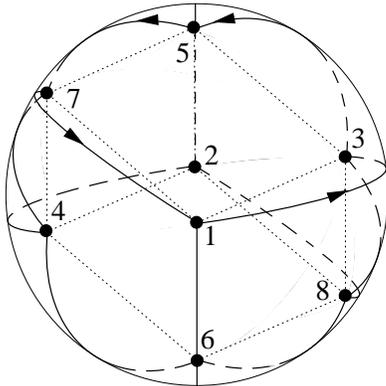}}
  \caption{The paths of the spin on the unit sphere 
between the easy positions of the field. 
The case of ${\cal C}({\bf O},3)$ configuration. }
  \label{fig:4a}
\end{figure}

In the case of ${\cal C}({\bf O},3)$ configuration, the minima are 
located at the vertices of a cube inscribed into the unit sphere: 
$\sin\theta=\sqrt{2/3},\;\sin(2\phi)=0$, where $\theta$ and $\phi$ 
are the spherical coordinates of the minima (see Fig.~\ref{fig:4a}). 
The tunneling trajectories divide the surface of the sphere 
into six congruent plaquettes; each subtends a solid angle of $2\pi/3$.
Five independent equations~(\ref{plaquette}) fix the tunneling 
phase shifts and the Hamiltonian up to an arbitrary gauge transformation. 
The eight eigenvalues are~\cite{cube}:

\bea
\label{8exact}
E_k^{\pm}&=&\pm2w\xi(\pi(J+3k)),\,\,k=0,\,1,\,2,\,3, \\
\xi(x)&=&\left(3+2\cos x\cos\frac{2x}{3}+
4\cos\frac{x}{2}\cos\frac{x}{3}\varrho(x)
\right)^{\frac12}, \nonumber \\
\varrho(x)&=&(4\sin^2\frac{x}{2}\sin^2\frac{x}{3}+1)^{\frac12}. \nonumber 
\eea

The ordered eigenvalues are presented in Table~\ref{tab:10} for the 
non-equivalent $J$'s ($w>0$). 
Analysis of the magnetic response is quite straightforward as 
well (see Appendix~\ref{appendix:B}); 
the magnetic susceptibilities of the classes are given in the 
last column of Table~\ref{tab:10}. 

\begin{table}
\caption{The spectra and the low temperature magnetic susceptibilities 
of ${\cal C}({\bf O},3)$ ($\beta=1/(k_{\rm B}T)$, 
a common factor of $(gJ\mu_{\rm B})^2$ is omitted).}
\begin{tabular}{lll}
$J$ & Eigenvalues(degeneracies) &Susceptibility\\ \hline
0 & $-3w\,(1)$, $-w\,(3)$, $w\,(3)$, $3w\,(1)$  & $1/(3w)$\\
1 & $-2w\,(3)$, $0\,(2)$, $2w\,(3)$            & $\beta/6$ \\
1/2 & $-\sqrt6w\,(2)$, $0\,(4)$, $\sqrt6w\,(2)$ & $\beta/9$\\
3/2 & $-\sqrt3w\,(4)$, $\sqrt3w\,(4)$ & $2\beta/9$ \\
\end{tabular}
\label{tab:10}
\end{table}

Consideration of ${\cal C}({\bf O},2)$ will be postponed 
till Section~\ref{section:Random}. 

\subsubsection{Multiple tunneling path regime.}
\label{mulitple-tunneling}

\begin{figure}
  \centerline{
\epsffile{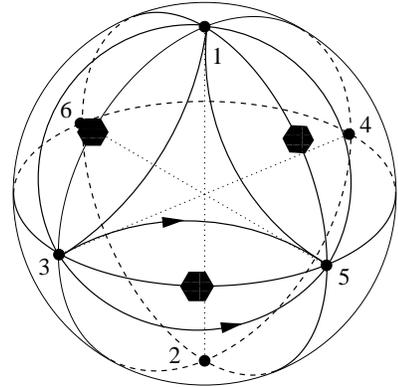}}
  \caption{Tunneling trajectories of the spin (double paths). 
The 6-fold configurations of ${\bf O}$. The hexagons show the 
locations of the maxima of the CEF potential.}
  \label{fig:3a}
\end{figure}

\begin{figure}
  \centerline{
\epsffile{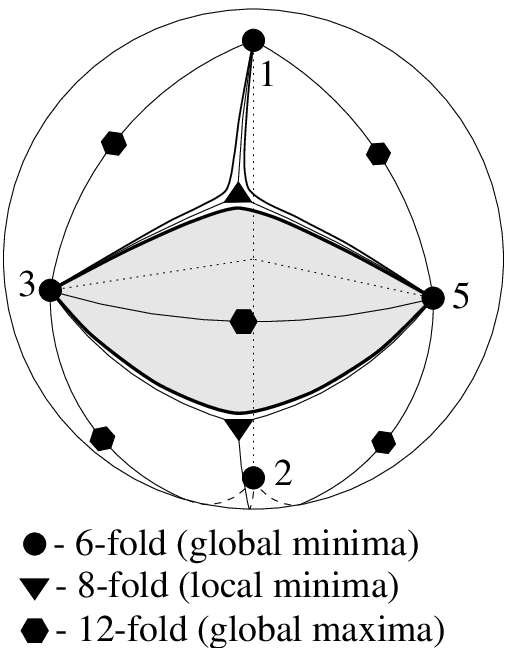}}
  \caption{Tunneling trajectories of the spin (double paths). 
The 6-fold configurations of ${\bf O}$. 
The tunneling trajectories pass 
closely to the local minima (locations of the 8-fold configuration).}
  \label{fig:3b}
\end{figure}

In ${\cal C}({\bf O},4)$ configuration, 
a tunneling trajectory connecting two minima, {\em e.g.}, minima 3 and 5 on 
Fig.~\ref{fig:1}, is not necessarily a geodesics on the sphere. 
For example, if the mid-point of the geodesics connecting minima 
3 and 5 is a maximum of the CEF potential then the tunneling trajectory 
connecting the minima will split in two paths: one deviating towards 
the ``north'' pole (minimum 1) and the other towards the ``south'' pole 
(minimum 2) as it is shown on Fig.~\ref{fig:3a}. One path is a mirror 
copy of the other with respect to the ``equatorial'' plane. 
Thus, the absolute values of the tunneling amplitudes corresponding 
to the two trajectories ($|w|$) are identical. To find the compound 
tunneling amplitude we assume that one of the trajectories, 
{\em e.g.}, the one connecting minima 3 and 5, and located in 
the ``south'' hemisphere, has the phase $\varphi_1$: 
$w_1=|w|\exp(i\varphi_1)$. Then, due to the Berry connection, 
the other amplitude must be $w_2=|w|\exp(i(\varphi_1-J\Omega))$, 
where $\Omega$ is the solid angle subtended by the two trajectories. 
The effective amplitude is: 
\be
w_e=w_1+w_2=2|w|e^{i(\varphi_1-J\Omega/2)}\cos(J\Omega/2). 
\label{double-trajectory}
\ee
Interesting conclusions can be derived from formula~(\ref{double-trajectory}). 
Firstly, the splitting of the trajectories does not change the connectivity 
matrix of the configuration it just modifies the multiplier of 
Hamiltonian~(\ref{Ham_O_6}) and all results obtained for configuration 
${\cal C}({\bf O},4)$ hold true. Secondly, the spectrum may be  
an oscillating function of $J$ or, if one would be able to 
vary parameters in such a way that $\Omega$ changes from its 
maximum value to zero, several oscillations of the spectrum could be 
observed as well. 
To estimate the number of oscillations we use the fact that: 
different tunneling trajectories emanated from a site and ending at some 
other site(s) do not intersect at intermediate points 
(they can only intersect at the end points). 
Then we can state that the maximal possible deviation of the trajectories 
from the spherical geodesics connecting the positions of 
${\cal C}({\bf O},4)$ configuration 
is reached when the trajectories pass along the spherical geodesics 
connecting the geometrically closest positions of 
${\cal C}({\bf O},3)$ and ${\cal C}({\bf O},4)$ configurations. 
Fig.~\ref{fig:3b} depicts this 
situation: the two tunneling trajectories connecting the 6-fold global minima 
3 and 5 (filled circles) are passing very closely to the 8-fold local minima 
(filled triangles), thus, ``avoiding'' the 12-fold global maxima 
(filled hexagons). 
The solid angle enclosed by the two trajectories 
(shaded area on Fig.~\ref{fig:3b}) 
varies in the range $0\leq\Omega<\pi/3$. Upon such a variation of $\Omega$ 
the spectrum will make $J/12$ full oscillations.  

\begin{figure}
  \centerline{
\epsffile{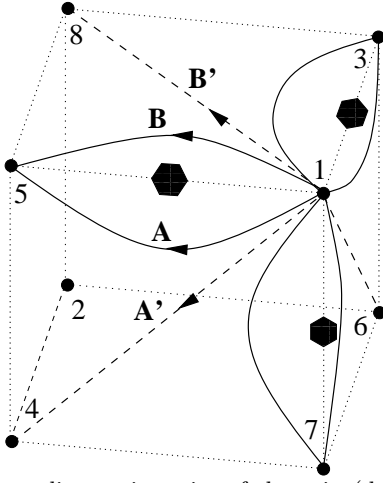}}
  \caption{Tunneling trajectories of the spin (double paths). 
The 8-fold configurations of ${\bf O}$. The hexagons show the 
locations of the maxima of the CEF potential.}
  \label{fig:3c}
\end{figure}

Next we analyze the multiple tunneling trajectories 
of ${\cal C}({\bf O},3)$. 
Fig.~\ref{fig:3c} depicts the splitting of the trajectory connecting 
minima 1 and 5 (solid curves A and B). 
The situation is similar to that of ${\cal C}({\bf O},4)$ configuration 
(the oscillations take place and their maximal number is $J/12$) 
except one subtle point: when a trajectory deviates strongly 
from the geodesics it approaches the trajectory connecting a 
next-nearest-neighbor (dashed lines on Fig.~\ref{fig:3c}), 
{\em e.g.}, lines A' and B' which connect 1 with 4 and 8 respectively. 
This is a very drastic change in the tunneling regime which leads 
to a change of the connectivity matrix. 

To calculate the spectrum we assume that the absolute values 
of the single tunneling amplitudes to the nearest- and 
next-nearest-neighbor sites are the same $w$. However, 
the effective amplitude for the nearest-neighbor tunneling 
is $2w\cos(J\Omega/2)$ due to the double trajectories. 
The elementary plaquette, in this case, is a triangle covering 
the solid angle of $\pi/3$, {\em e.g}, triangle 1-5-8-1 
on Fig.~\ref{fig:3c}. 
In this case, plaquettes cover the sphere twice. 
Then, the periodicity of the spectra is given by $s=4\pi/(\pi/3)$, which 
is a half of the total number of the elementary plaquettes.
Application of the symmetry arguments 
given at the beginning of this section leads to the following 
properties of the spectrum: the periodicity of the spectral 
behavior is $J=s/2=6$, $|J+6n|,\,n\in{\bf Z}$ is equivalent to $J$, 
the spectrum of $J+3$ is the inverted spectrum of $J$. 
The results of the diagonalization are summarized 
in Table~\ref{tab:10a}. 

\begin{table}
\caption{The spectra of ${\cal C}({\bf O},3)$,  
the region of the multiple tunneling path regime [$x=2\cos(J\Omega/2)$]; 
all eigenvalues are multiples of $w$.}
\begin{tabular}{lll}
$J$ & Eigenvalues(degeneracies) \\ \hline
0 & $-3(1-x)\,(1)$, $-(1+x)\,(3)$, $(1-x)\,(3)$, $3(1+x)\,(1)$  \\
1 & $-2(1-x/2)\,(3)$, $-3x\,(2)$, $2(1+x/2)\,(3)$            \\
2 & $-2(1+x/2)\,(3)$, $3x\,(2)$, $2(1-x/2)\,(3)$            \\
3 & $-3(1+x)\,(1)$, $-(1-x)\,(3)$, $(1+x)\,(3)$, $3(1-x)\,(1)$  \\
1/2 & $-(\sqrt6-x\sqrt3)\,(2)$, $-x\sqrt3\,(4)$, $(\sqrt6+x\sqrt3)\,(2)$ \\
3/2 & $-\sqrt{3(1+x^2)}\,(4)$, $\sqrt{3(1+x^2)}\,(4)$ \\
5/2 & $-(\sqrt6+x\sqrt3)\,(2)$, $x\sqrt3\,(4)$, $(\sqrt6-x\sqrt3)\,(2)$ \\
\end{tabular}
\label{tab:10a}
\end{table}

\subsection{Spectra of the ${\bf Y}$ configurations}
\label{spectrum-Y}

The analysis of the configurations of ${\bf Y}$ group is 
tedious, though similar to that for ${\bf O}$ group. 
We present only the results of the analysis here. 
Table~\ref{tab:11} contains the spectra and the 
low temperature susceptibilities of 
${\cal C}({\bf Y},5)$ configuration (the energies are multiples of $w$). 
Tables~\ref{tab:12},~\ref{tab:13} contain the spectra and the 
low temperature susceptibilities of 
${\cal C}({\bf Y},3)$ configuration respectively. 

\begin{center}
\begin{table}
\caption{The spectra and the low temperature magnetic susceptibilities 
of ${\cal C}({\bf Y},5)$; all eigenvalues are multiples of $w$ 
and all susceptibilities are multiples of $(gJ\mu_{\rm B})^2$ 
[$\beta=1/(k_{\rm B}T)$, $c_1=\cos(\pi/10)$, and $c_3=\cos(3\pi/10)$].}
\begin{tabular}{lll}
$J$ & Eigenvalues(degeneracies) &Susceptibility\\ \hline
$0$       & $-\sqrt5\,(3)$,  $-1\,(5)$,  $\sqrt5\,(3)$,  $5\,(1)$ &
$(1+\sqrt5)/(6w)$ \\
$1$   & $-\sqrt5\,(4)$, $(\sqrt5-3)/2\,(5)$, $(5+\sqrt5)/2\,(3)$ & $\beta/9$ \\
$2$   & $-\sqrt5\,(4)$, $(\sqrt5-5)/2\,(3)$, $(\sqrt5+3)/2\,(5)$ & $\beta/9$ \\
$3$  & $-(\sqrt5+3)/2\,(5)$, $(5-\sqrt5)/2\,(3)$, $\sqrt5\,(4)$ & $2\beta/9$\\
$4$   & $-(5+\sqrt5)/2\,(3)$, $(3-\sqrt5)/2\,(5)$, $\sqrt5\,(4)$ & 
$\beta/6$ \\ 
$5$  & $-5\,(1)$, $-\sqrt5\,(3)$, $1\,(5)$, $\sqrt5\,(3)$ & 
$(5+\sqrt5)/(30w)$ \\
$\frac12$ & $-2 c_1\,(6)$, $(3-\sqrt5) c_1\,(4)$,  
$2\sqrt5c_1\,(2)$ & $\beta/5$ \\
$\frac32$ & $-2\sqrt5 c_3\,(2)$, $-2 c_3\,(6)$,  
$(3+\sqrt5)c_3\,(4)$  
& $(5+\sqrt5) c_1/(15w)$ \\
$\frac52$ & $-\sqrt5\,(6)$, $\sqrt5\,(6)$ & $\beta(5+\sqrt5)/30$ \\
$\frac72$ & $-(3+\sqrt5) c_3\,(4)$, $2c_3\,(6)$, 
$2\sqrt5 c_3\,(2)$, & $\beta/5$ \\
$\frac92$ & $-2\sqrt5 c_1\,(2)$, 
$(\sqrt5-3) c_1\,(4)$, $2c_1\,(6)$ & $\beta/9$ \\
\end{tabular}
\label{tab:11}
\end{table}
\end{center}

\begin{center}
\begin{table}
\caption{The spectra of ${\cal C}({\bf Y},3)$; 
all eigenvalues are multiples of $w$.}
\begin{tabular}{ll}
$J$ & Eigenvalues(degeneracies) 
\\ \hline
$0$ & $-\sqrt5\,(3)$, $-2\,(4)$,  $0\,(4)$, $1\,(5)$,  
$\sqrt5\,(3)$, $3\,(1)$  \\
$1$ & 
\begin{tabular}{l}
$-(1+\sqrt{13})/2\,(5)$, $-1\,(4)$, 
$(3-\sqrt5)/2\,(3)$, \\
$(-1+\sqrt{13})/2\,(5)$, 
$(3+\sqrt5)/2\,(3)$  
\end{tabular} \\
$2$  & 
\begin{tabular}{l}
$-(3+\sqrt5)/2\,(3)$, $(1-\sqrt{13})/2\,(5)$, 
$(3-\sqrt5)/2\,(3)$, \\ 
$1\,(4)$, $(1+\sqrt{13})/2\,(5)$  
\end{tabular} \\
$3$ & $-3\,(1)$, $-\sqrt5\,(3)$, $-1\,(5)$, $0\,(4)$, 
$2\,(4)$, $\sqrt5\,(3)$ \\
$1/2$ & 
\begin{tabular}{l}
$-(\sqrt3+\sqrt7)/2\,(6)$, $\sqrt3(1-\sqrt5)/2\,(2)$, \\ 
$(-\sqrt3+\sqrt7)/2\,(6)$, $\sqrt3\,(4)$, 
$\sqrt3(1+\sqrt5)/2\,(2)$ \\
\end{tabular}  \\
$3/2$ & $-\sqrt6\,(4)$, $-1\,(6)$, $1\,(6)$, $\sqrt6\,(4)$    \\
$5/2$ &  
\begin{tabular}{l}
$-\sqrt3(1+\sqrt5)/2\,(2)$, $-\sqrt3\,(4)$, 
$(\sqrt3-\sqrt7)/2\,(6)$ \\ 
$\sqrt3(-1+\sqrt5)/2\,(2)$, 
$(\sqrt3+\sqrt7)/2\,(6)$  \\
\end{tabular} \\
\end{tabular}
\label{tab:12}
\end{table}
\end{center}

\begin{center}
\begin{table}
\caption{Low Temperature magnetic susceptibilities 
of ${\cal C}({\bf Y},3)$ 
[a common factor of $(gJ\mu_{\rm B})^2$ is omitted, $\beta=1/(k_{\rm B}T)$].}
\begin{tabular}{cc}
$J$ & Susceptibility \\
\hline 
0 & $(7/6+11\sqrt5/18)/w$ \\
1 & $\beta(12\sqrt5+37+\sqrt{13}(3\sqrt5+4))/468$ \\
2 & $\beta/6$ \\
3 & $(\sqrt5+3)/(6w)$ \\
1/2 & $\beta(10\sqrt5+83+\sqrt{21}(5\sqrt5-2))/630$ \\
3/2 & $\beta(4\sqrt5+9)/90$ \\
5/2 & $\beta/9$ \\
\end{tabular}
\label{tab:13}
\end{table}
\end{center}

The multiple tunneling path regime is present in configurations 
${\cal C}({\bf Y},5)$ and ${\cal C}({\bf Y},3)$ as well. 
Its analysis is similar to that of configurations 
${\cal C}({\bf O},4)$ and ${\cal C}({\bf O},3)$. 
We present here its summary only: 
The regions of existence of configurations  
${\cal C}({\bf Y},5)$ and ${\cal C}({\bf Y},3)$ 
in the parameter space of the CEF are divided 
into two parts for each configuration. One part corresponds to 
the single tunneling path regime. 
The above theory is valid in this region. 
The other part is of the multiple tunneling path regime. 
The spectra are oscillating functions of $J$ in this region 
since $w\sim\cos(J\Omega/2)$, $0\leq\Omega< 2\pi/15$. 
Upon full monotonic variation of $\Omega$, the spectra makes 
$\approx J/30$ oscillations for the both configurations. 
The spectra of ${\cal C}({\bf Y},5)$ given in Table~\ref{tab:11} 
holds valid for the both regimes. For ${\cal C}({\bf Y},3)$ configuration, 
in a range of parameters the proximity of the minima 
positions may be altered: each minimum position 
(vertex of the dodecahedron where some three faces intersect) 
should be geometrically connected not just to the three nearest-neighbors 
but also to the six next-nearest-neighbors.

\section{Random energy levels.}
\label{section:Random}

\begin{figure}
  \centerline{
\epsffile{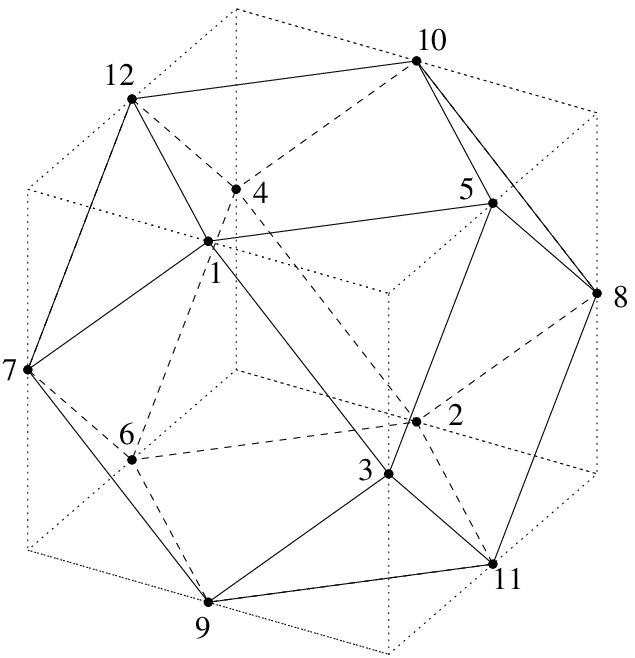}}
  \caption{Minima distribution of  
the 12-fold configurations of ${\bf O}$.}
  \label{fig:5}
\end{figure}

For all configurations considered in previous section, 
the spectra were simple periodic functions of $J$, which 
was due to the fact that a rational number 
of flux quanta ($\Phi=J\Omega(c)=2\pi JP/Q$) passes through each 
plaquette. This is not the case for more complex configurations 
such as ${\cal C}(G,2)$, $G={\bf O,\,Y}$. 
In Fig.~\ref{fig:5} we present the spatial distribution 
of minima of ${\cal C}({\bf O},2)$ configuration. 
The segments connecting the minima are not real tunneling 
trajectories but rather guidelines. The tunneling paths 
may deviate strongly from the geodesics connecting corresponding 
minima both to the locations 
of the 6-fold (centers of the cube faces) and 8-fold 
(vertices of the cube) configurations' positions. 
The exact form of the paths depends on the CEF constants, 
{\em e.g.}, for the simplest Hamiltonian~(\ref{hamilton2}), 
where configuration ${\cal C}({\bf O},2)$ is realized, 
it is a function of ratio $b/a$. Instead of studying non-universal 
tunneling trajectories, we introduce a parameter 
$\alpha$ ($0<\alpha<2\pi/3$): 
the solid angle subtended by a square-like contour. 
The solid angle subtended 
by a triangle-like circuit is $\pi/2-3\alpha/4$. 
A knowledge of this parameter together with $w$ is sufficient to 
define the spectra of the 12-fold configuration. 
Since $\alpha$ may be an irrational multiple of $\pi$, 
the spectra as a function of $J$ is not expected to be a 
finite set of values, but a fractal set. 
The spectra of the 12-fold configuration are 
given in Table~\ref{tab:14}. The spectra undergo 
$J/12$ oscillations upon a monotonic variation of $0<\alpha<2\pi/3$
for a given value of spin $J$. 

\begin{figure}
  \centerline{
\epsffile{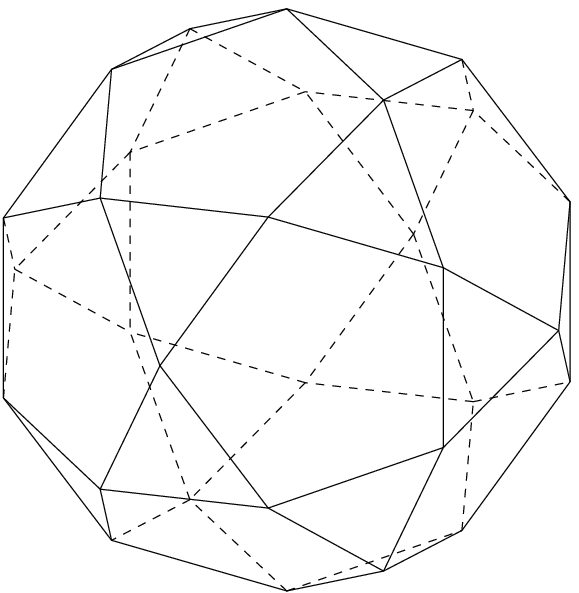}
}  
\caption{Minima distribution of  
the 30-fold configurations of ${\bf Y}$.}
  \label{fig:6}
\end{figure}

Configuration ${\cal C}({\bf Y},2)$ is even more complex than 
${\cal C}({\bf O},2)$. Its minima directions correspond to 
the midpoints of the icosahedron edges (see Fig.~\ref{fig:6}). 
The parameter $\alpha$ ($0\leq\alpha<\pi/3$) 
here corresponds to the solid angle 
subtended by a pentagon-like contour. The spectra undergo 
$J/30$ oscillations upon a monotonic variation of $0<\alpha<\pi/3$.
The spectra of the 30-fold configuration for odd values of $J$ 
are given in Table~\ref{tab:15}. 

The spectra described in this section have features of randomness. 
Indeed, the function $\{\alpha J\}$ (fractional part of $\alpha J$) 
with an irrational $\alpha$ is known as a generator of random numbers. 
Thus, the ratios of the transition frequencies for configurations 
${\cal C}(G,2)$, $G={\bf O,\,Y}$ vary in an uncontrollable way 
when large $J$ changes by 1. This behavior differs dramatically from that 
for other cubic and icosahedral configurations which display permanent 
ratios of the frequencies for a fixed configuration. Thus, the configurations 
${\cal C}(G,2)$, $G={\bf O,\,Y}$ realize the chaotic spectra of deterministic 
systems. This situation is well-known, {\em e.g.}, for the Hydrogen atom 
in a uniform magnetic field~\cite{hydrogen}. 
The peculiarity of our problem is that it 
displays chaos in a finite set of numbers (12 or 30) and that the 
chaotic behavior can be found analytically. Another special feature 
of our system is that stochasticity in it is combined with deterministic 
multiplicity distribution. For example, in the case of the 
${\cal C}({\bf O},2)$ configuration the 12 levels are divided into 
submultiplets given in Table~\ref{tab:2}, independently on $\alpha$. 
However, their mutual arrangement is unpredictable. 

For a two parametric Hamiltonian, {\em e.g.}, 
Hamiltonian~(\ref{hamilton2}) for the octahedron group, 
the configurations ${\cal C}(G,2)$, $G={\bf O,\,Y}$ correspond to 
the single tunneling regime. The multiple tunneling regime 
may occur if the invariants of higher orders are included. 

\begin{center}
\begin{table}
\caption{Spectra of ${\cal C}({\bf O},2)$. 
All eigenvalues are multiples of $w$; $x=J(\alpha+2\pi)/4$.}
\begin{tabular}{l|l}
Integer $J$  & 
Half-integer $J$  \\
\hline
Energy(Degeneracy) & Energy(Degeneracy)\\
\hline
$4\cos x$ (1) & $2(\cos x\pm\sqrt2\sin x)$ (2,2) \\
$-2\cos x$ (2) & $-\cos x\pm\sqrt{2+\cos^2x}$ (4,4) \\
$2\cos x$ (3) & \\
$-\cos x\pm\sqrt{8-7\cos^2x}$ (3,3) & \\
\end{tabular}
\label{tab:14}
\end{table}
\end{center}

\begin{center}
\begin{table}
\caption{Spectra of ${\cal C}({\bf Y},2)$ for odd values of $J$. 
All eigenvalues are multiples of $w$, $x=\cos(J(\alpha+3\pi)/5)$.}
\begin{tabular}{l}
Energy(Degeneracy) \\
\hline
$\pm1+2x$ (4,4) \\
$-x\pm\sqrt{4-3x^2}$ (5,5) \\
$(1+\sqrt5)(-x\pm\sqrt{4+(5-4\sqrt5)x^2}\,\,\,)/2$ (3,3)  \\
$(1-\sqrt5)(-x\pm\sqrt{4+(5+4\sqrt5)x^2}\,\,\,)/2$ (3,3)  \\
\end{tabular}
\label{tab:15}
\end{table}
\end{center}

In the presence of infinitely small magnetic field the ground-state 
of configuration ${\cal C}({\bf O},2)$ acquires either a finite magnetic
moment or a finite susceptibility. We analyzed this problem for the 
field directed along one of the fourth order axes and $w>0$. 
Then the finite magnetic moment 
$2gJ\mu_{\rm B}|\sin x|/(2(8-7\cos^2 x))^{1/2}$ is acquired at 
$\cos x > -1/2$ ($x=J(\alpha+2\pi)/4$), 
otherwise the finite magnetic susceptibility 
$\chi=-(gJ\mu_{\rm B})^2/(3w\cos x)$ occurs for integer $J$. 
For half-integer $J$, the magnetic moment $gJ\mu_{\rm B}/3$ is acquired 
at $\cos x < \cos(3\pi/8)$, otherwise this value of the moment 
is multiplied by a factor: 
$$
\left(\frac{c^2+5+3c\sqrt{2+c^2}+2\sqrt2|\sin(x)|(3c+\sqrt{2+c^2})}{2(2+c^2)}
\right)^{\frac12}, 
$$
where $c=\cos(x)$. 
Note the random character of these values. 
\section{Numerical analysis. The case of the cubic CEF.}
\label{section:Numerical}

The main obstacle to a reliable numerical analysis of the problem 
is the fact that nobody knows how the Hamiltonian looks like. 
The case of the rare-earth ions with large total angular momenta 
interacting with the CEF represents an exception. 
Only the orbital part ${\bf L}$ of the total angular momentum of 
a single magnetic electron interacts with the crystalline field. 
All terms, in the expansion of the crystalline potential 
with the degree larger than $2l$, where $l$ is the orbital quantum 
number of the single magnetic electron, 
vanish~\cite{Hutch}, thus, 
simplifying the analysis. 
For the $4f$-group electrons with $l=3$, this gives the highest 
non-vanishing terms of the sixth order. Considering a CEF of 
a particular symmetry group brings further simplification, 
{\em e.g.}, in the cubic CEF, there are only two independent 
invariants of the sixth order and one of the fours order. 
The two of the sixth order are combined in one invariant 
(see Eq.~(\ref{hamilton2})) for a real interaction which is 
the Coulomb interaction between the charge carriers. 

It has been shown in Section~\ref{section:Quasiclassics} that 
Hamiltonian~(\ref{hamilton2}) has configurations 
${\cal C}({\bf O},4)$, ${\cal C}({\bf O},3)$, and 
${\cal C}({\bf O},2)$ as sets of its classical extrema. 
In this meaning it is rather general. 
Therefore, we apply numerical analysis to this Hamiltonian 
in a wide range of $J$'s. 
It means that we diagonalize numerically $(2J+1)\times(2J+1)$ matrix 
for one-parametric set of Hamiltonians~(\ref{Ohamilton2}). The 
choice of this Hamiltonian is partly justified by the above consideration. 
Our purpose is to find numerically 
what $J$ can be considered as large, {\em i.e}, starting from 
what $J$ our theory gives satisfactory description. 
The second important problem is the crossover behavior 
of the spectrum near configuration boundaries described in 
Section~\ref{section:Quasiclassics}.

First numerical studies of the crystal field effects on angular 
momenta were performed in the early 60's by Lea, Leask, and 
Wolf~\cite{LLW}. These authors studied a cubic crystal field 
Hamiltonian similar to~(\ref{hamilton2}). Their main interest was 
how the  angular momentum degeneracy of $f$-electrons is lifted. 
For this purpose it was enough to consider values of $J$ spanned from 3 to 8. 
Refraining ourself from this limitations, we study numerically 
the Hamiltonian consisting of terms of the fourth and sixth order 
for an arbitrary value of $J$. 
However, we use a different parameterization than that used in~\cite{LLW} 
for the same Hamiltonian:
\be
H_2^{\bf O}=-\frac{\cos(\phi)O_4^0}{(J(J+1))^2}-
\frac{5}{14}\frac{\sin(\phi)O_6^0}{(J(J+1))^3}, 
\label{Ohamilton2}
\ee
where $O_4^0$ and $O_6^0$ are Stevens' operator 
equivalents~\cite{Stevens,Hutch}, and $\phi$ is a parameter taking 
values in the interval $[-\pi,\pi]$. Our parameterization corresponds 
to a unit circle on the phase diagram of Hamiltonian~(\ref{hamilton2}) 
(see Fig.~\ref{fig:3}): $a=\cos(\phi),\,b=\sin(\phi)$, 
whereas that chosen in~\cite{LLW} corresponds to 
the square: $a=x,\,14b/5=\pm(1-|x|);\,-1\leq x\leq 1$. 
The coefficient of $5/14$ reflects 
the difference between our invariant of the sixth order 
in~(\ref{hamilton2}) and the commonly used 
Stevens' operator equivalent $O_6^0$. 

For relatively small values of spins $2J+1\sim N$, 
where $N$ is the number of extrema of the CEF, the quasi-classical 
description fails and the spectrum of Hamiltonian~(\ref{Ohamilton2}) 
does not follow the predicted dependence. 
However, for $J\approx10$, one can observe distinct regions of 
$\phi$ with high density of level crossing 
(these regions are distinctly seen in~\cite{LLW} for $J\geq6$). 
Upon an increase of $J$ these regions narrow down giving the points 
separating the 6-, 8-, and 12-fold configurations. 
Further increase of $J$ leads to a ``bunching'' of low energetic levels 
into the predicted groups (multiplets) of six, eight, or twelve. 

Not only the numbers of the levels in the multiplets, 
but also the ratios of the spacings between the 
levels inside the multiplets, the oscillations of the spectra in 
the regime of the multiple tunneling path, and the tunneling amplitude 
in the regime of a single tunneling path obey the predictions of our theory. 

\begin{figure}
  \centerline{
\epsffile{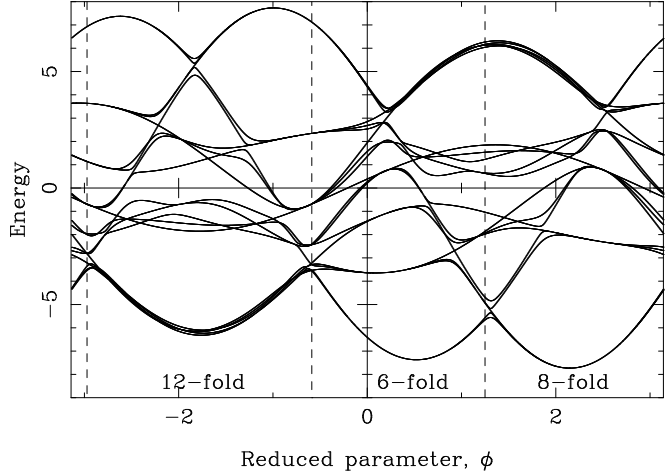}}
  \caption{The spectrum of Hamiltonian~(\ref{Ohamilton2}); $J=23$.}
  \label{fig:9a}
\end{figure}

\begin{figure}
  \centerline{
\epsffile{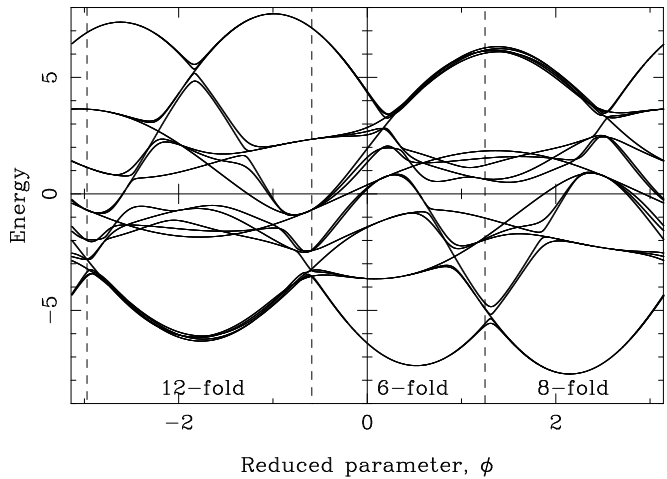}}
  \caption{The spectrum of Hamiltonian~(\ref{Ohamilton2}); $J=47/2$.}
  \label{fig:9b}
\end{figure}

\begin{figure}
  \centerline{
\epsffile{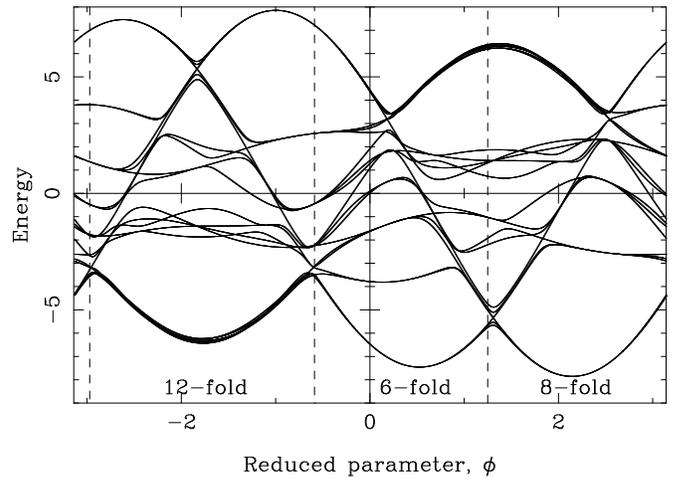}}
  \caption{The spectrum of Hamiltonian~(\ref{Ohamilton2}); $J=24$.}
  \label{fig:9c}
\end{figure}

For a demonstration we have chosen a set of close valued $J$'s: 
$J=23$, $47/2$, and $24$. 
Figs.~\ref{fig:9a}, \ref{fig:9b}, and \ref{fig:9c} 
are graphs of the spectra of 
Hamiltonian~(\ref{Ohamilton2}) for these values of $J$. 
The vertical dashed lines are the classical boundaries 
between the different configuration 
(see the diagram of Hamiltonian~(\ref{hamilton2}) Fig.~\ref{fig:3}). 
A small deviation of the dashed line separating 
the 6- and 8- fold configurations ($\phi_{6-8}=\arctan(3)$) 
towards the 6-fold one is due to the fact that, 
at $\phi=\phi_{6-8}$, the depth of the CEF potential in 
the minimum locations of the 6-fold configuration is equal to that of 
the 8-fold configuration. However, the intersection of the levels 
occurs when the ground-state energies coincide. 
See Appendix~\ref{appendix:O14} for details on this subject.

From the pictures one can clearly see the ``bunching'' 
of the highest and lowest energy 
levels into the predicted multiplets of 6, 8, and 12. 
The excited multiplets have the same structure which 
fails only in the vicinity of the boundaries between the configurations. 
The structure of the spectra given on 
Figs.~\ref{fig:9a}, \ref{fig:9b}, and \ref{fig:9c}
looks quite similar at this level of ``magnification''. 
To see the subtle details predicted in previous sections we 
should ``zoom in'' the pictures ``focusing'' on the ground multiplet.

\begin{figure}
  \centerline{
\epsffile{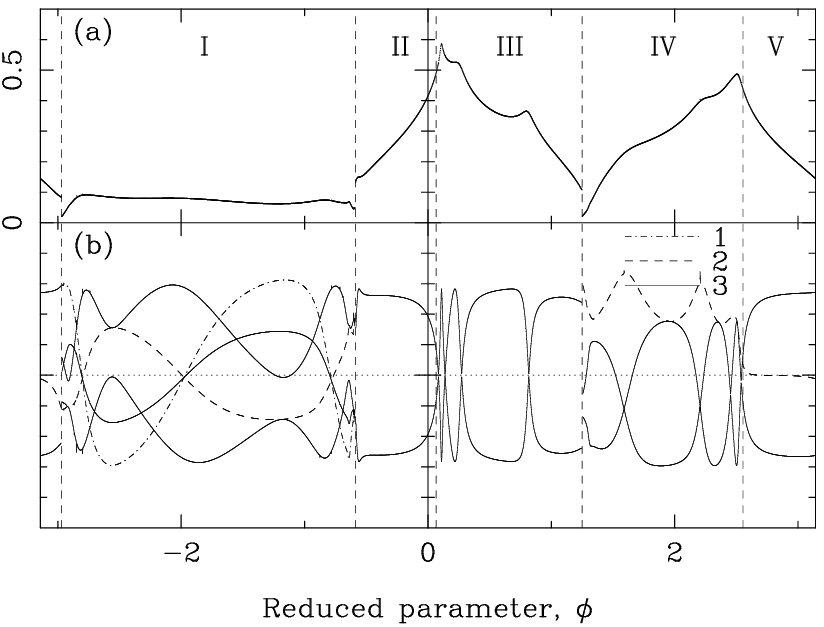}}
  \caption{(a) Graph of $\-ln(R)/J$, where $R$ is the 
rescaling factor applied to the ground multiplet; 
(b) Rescaled ground state multiplet of 
Hamiltonian~(\ref{Ohamilton2}) (the legend shows the 
degeneracies of the levels); $J=23$.}
  \label{fig:10a}
\end{figure}

\begin{figure}
  \centerline{
\epsffile{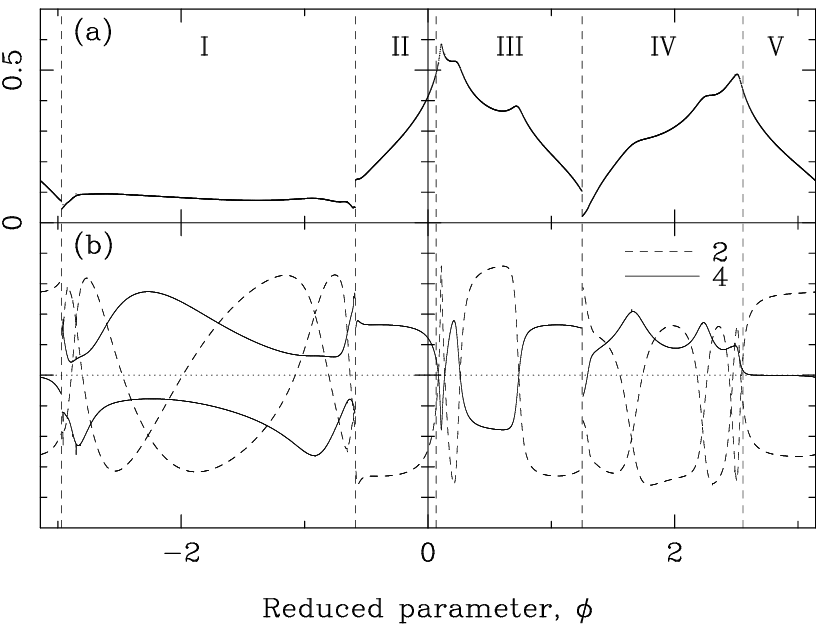}}
  \caption{(a) Graph of $\-ln(R)/J$, where $R$ is the 
rescaling factor applied to the ground multiplet; 
(b) Rescaled ground state multiplet of 
Hamiltonian~(\ref{Ohamilton2}) (the legend shows the 
degeneracies of the levels); $J=47/2$.}
  \label{fig:10b}
\end{figure}

\begin{figure}
  \centerline{
\epsffile{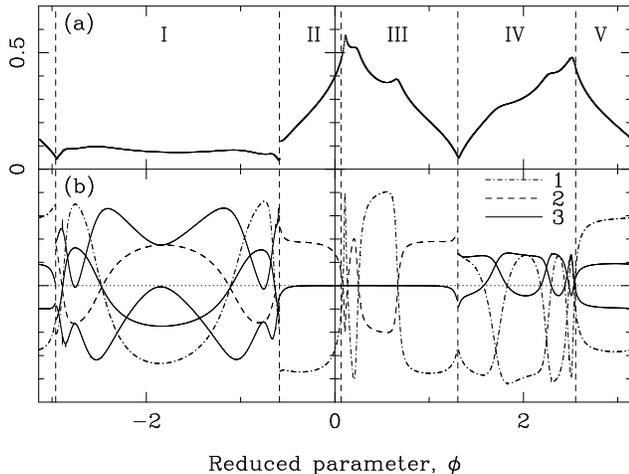}}
  \caption{(a) Graph of $\-ln(R)/J$, where $R$ is the 
rescaling factor applied to the ground multiplet; 
(b) Rescaled ground state multiplet of 
Hamiltonian~(\ref{Ohamilton2}) (the legend shows the 
degeneracies of the levels); $J=24$.}
  \label{fig:10c}
\end{figure}

Firstly, we shift the ``center of mass'' of the ground multiplet 
to zero (we are not interested in finding the single well 
localization energy). 
Secondly, we rescale the shifted levels, 
so that a ``visual'' comparison of the 
spacings between the levels can be done at different values of the reduced 
parameter $\phi$. 
The rescaling is necessary due to a large variation of 
$w\propto\exp(-Jc(\phi))$.
The calculations of the tunneling amplitude for 
${\cal C}({\bf O},4)$ configuration 
of Hamiltonian~\ref{hamilton2} (see Fig.~\ref{fig:2}) predict a variation 
of $w$ of order $10^6$ for $J\approx24$. 
The results of this program are shown on 
Figs.~\ref{fig:10a}, \ref{fig:10b}, and~\ref{fig:10c} for 
$J=23$, $47/2$, and $24$ respectively. The vertical dashed lines 
(the quasiclassical boundaries, see Fig.~\ref{fig:3})
separate not only the regions of different configuration numbers but 
also the regions of the single and multiple tunneling path regimes. 
The regions are enumerated by Roman numerals: 
I - ${\cal C}({\bf O},2)$ (single tunneling path regime only), 
II - ${\cal C}({\bf O},4)$ single, 
III - ${\cal C}({\bf O},4)$ multiple, 
IV - ${\cal C}({\bf O},3)$ multiple, and 
V - ${\cal C}({\bf O},3)$ single regimes respectively. 
Part (a) of each picture represents the plot of the rescaling factor 
which is proportional to $c(\phi)$; part (b) is the rescaled spectrum. 

All predictions of Sections~\ref{section:Quasiclassics}, 
\ref{section:Spectrum} and~\ref{section:Random} 
(the orderings of the levels, 
the ratios of the level spacings, the oscillations of 
the spectra in some regions, the numbers of the oscillations, and 
the dependence of the scaling parameter $R$) find confirmation here. 
The oscillations are not of the periodic form due to 
a non-trivial (but monotonic) dependencies $\alpha=\alpha(\phi)$ and 
$\Omega=\Omega(\phi)$. 

\begin{figure}
  \centerline{
\epsffile{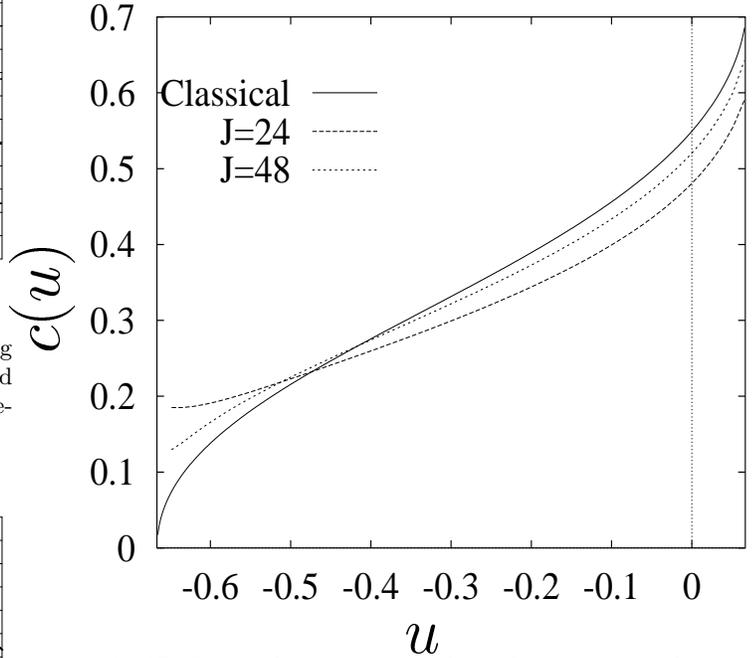}}
  \caption{Comparison of the quasi-classical and numerical tunneling amplitude 
exponent, $u=\tan(\phi)$.}
  \label{fig:11}
\end{figure}

A more precise value of $c(\phi)$ can be easily obtained 
from the single tunneling path part of the 
spectrum of the 6-fold configuration. Fig.~\ref{fig:11} 
compares the quasi-classical result found in 
Section~\ref{section:Quasiclassics} with the numerical calculations 
for $J=24$ and 48. 
The plot is $-\ln((E_1-E_0)/4)/J$ vs. $u=tan(\phi)$, where $E_0$ and $E_1$ 
are the energies of the ground and first excited states respectively. 
The difference $E_1-E_0$ is $4w$ according to predictions of 
Section~\ref{section:Spectrum}. 
A small discrepancy is due to the coefficient of 
the exponential $f(u)$ ($w=f(u)\exp(-Jc(u))$), whose contribution 
decreases $\propto 1/J$. From these data, we can estimate 
that the values of the coefficient $f(u)$ are in a range 0.1---3.0. 

All these facts strongly emphasize the validity of the developed 
quasi-classical description of the large spins from the theoretical point 
of view. 
Now questions arise: What is a possible experimental realization? 
What are the limitations of the theory when 
applied to the real systems?
We will elaborate these question in the next section.
\section{Experimental realization}
\label{section:Experiment}

\subsection{Feasible experimental systems}
\label{systems}

The experimental observation of the predicted effects 
can be done on any system with large values of 
the angular momentum such as rare-earth ions, magnetic clusters, 
or nuclei. The main question is whether the value of $J$ is 
large enough for a given configuration of the external field. 

For configuration with small number of minima (2- and 4-fold 
configuration), $J\approx8$ satisfies the quasi-classical 
requirement. Such values of $J$ are available, {\em e.g.}, 
in rare-earth ions: Dy$^{+3}$, Ho$^{+3}$, or Er$^{+3}$. 
An example of compounds with the tetragonal symmetry, 
where the 4-fold configuration is realized, is 
RENi$_2$B$_2$C, RE stands for a rare-earth magnetic element. 
To suppress the influence of the interaction between the 
magnetic moments the magnetic ions should be diluted with 
similar but non-magnetic ones such as La$^{+3}$, Lu$^{+3}$, 
or Y$^{+3}$. The CEF effects for this family of compounds 
were studied in single crystals of Lu$_{1-x}$Ho$_x$Ni$_2$B$_2$C 
by Cho {\em et al.}~\cite{Cho}. The calculated CEF level scheme 
given there shows that the ground state quadruplet is well 
separated from other excited states and corresponds 
to the multiplet of ${\cal C}({\bf D}_4,2)$ configuration with 
$J=0 \!\pmod{2}$ and $w\approx2$K. 

Another family of rare-earth compounds, RESb, offers the 
cubic environment. However, it is questionable whether 
the quasi-classical requirement is satisfied since 
even for the highest values of the angular moment 
($J=8$ for Ho$^{+3}$) the multiplicity $2J+1=17$ is not so 
large comparatively to the lowest dimension of the cubic 
configurations $N=6$. The numerical calculations performed in the 
previous section indicate that only for $J\geq12$ there are 
regions of parameter $\phi$ where the 6- and 8-fold configurations 
are well defined. To obtain the 12-fold configuration, 
in the framework of Hamiltonian~(\ref{Ohamilton2}), the value 
of $J$ should be increased to about 24. 

Magnetic clusters and molecules offer systems with 
very large total spins and a variety of symmetries. 
Theoretical calculations~\cite{Jongh} indicate that clusters of 
13 atoms of transition metals such as Fe, Pd, and Rh may have 
cubic symmetry and total magnetic moment of the order of 
$\mu_{\rm B}$ per atom.
Gadolinium clusters Gd$_n$ ($n=11-92$)~\cite{Gd} exhibit 
large magnetic moments of 0.5-3.0$\mu_{B}$ per atom 
(which is below the bulk value of 7.55$\mu_{B}$ but, 
still offers a large value of the total cluster spin) 
with behaviors ranging from tight locking to the lattice 
by crystal anisotropy to superparamagnetism (almost free moment). 

Large spins were also observed in artificially grown magnetic dots 
used for observation of the magnetic tunneling~\cite{Barbara}. 
So far, these systems belonged to the lowest symmetry class. 
it is rather tempting to create environment of higher symmetry and 
to use smaller magnetic dots like the ones used by I.~Schuller 
and {\em et al}~\cite{Schuller} to observe the effects predicted
by our theory. 

\subsection{Practicable experiments. Magnetic measurements.}
\label{experiments}

The experimental consequences of the difference among 
the configurations and the spin values 
can be observed with many experiments. 
To name a few these are 
measurements of the spin magnetic moment and magnetic susceptibility, 
relaxation of the magnetization, 
electron paramagnetic resonance (EPR), and 
nuclear magnetic resonance (NMR). 

First we discuss measurements of the magnetic susceptibility.
The magnetic susceptibility follows the Curie law for temperatures higher than 
the characteristic splitting of the ground-state multiplet 
$Tk_{\rm B}>w$ (see Appendix~\ref{appendix:B}). 
Thus, it is $(g\mu_{\rm B}J)^2/(k_{\rm B}T)$ for 
1-dimensional configurations, 
{\em i.e.}, ${\cal C}({\bf D}_N,N)$, $N=2,\,4,\,6$, 
$(g\mu_{\rm B}J)^2/(2k_{\rm B}T)$ for 2-dimensional ones, 
{\em i.e.}, ${\cal C}({\bf D}_N,2)$, $N=4,\,6$, and 
$(g\mu_{\rm B}J)^2/(3k_{\rm B}T)$ for the 3-dimensional ones, 
{\em i.e.}, for the rest of configurations considered in this work.
For temperatures lower than the characteristic splitting $Tk_{\rm B}<w$, 
the Curie dependence is no longer universal. 
The non-magnetic classes, that is those without magnetic moment in the 
ground state, have their magnetic susceptibility saturated to some constant 
at $T\rightarrow 0$, 
whereas, the magnetic ones (with a non-zero magnetization 
in the ground state) still obey the Curie-like behavior. 
Both the saturation values and the slopes of the 
Curie-like curves depend upon the configuration of the symmetry 
group as well as upon the equivalence class of the spin. 
For example, the 2-fold configurations 
${\cal C}({\bf D}_N,N)$ are non-magnetic for $J\cong N/2$, $N=2,\,4,\,6$; 
the saturation values of the magnetic susceptibility are 
$\chi(T)=(g\mu_{\rm B}J)^2/|E(0)|$, 
where $E(0)$ are the corresponding eigenvalues for zero magnetic field 
(see Eqs.~(\ref{D-2-fold})). The magnetic classes of these configurations, 
{\em i.e.}, the ones with $J\not\cong N/2$, 
have the same Curie-like dependence: 
$\chi(T)=(g\mu_{\rm B}J)^2/(k_{\rm B}T)$.

\begin{figure}
  \centerline{
\epsffile{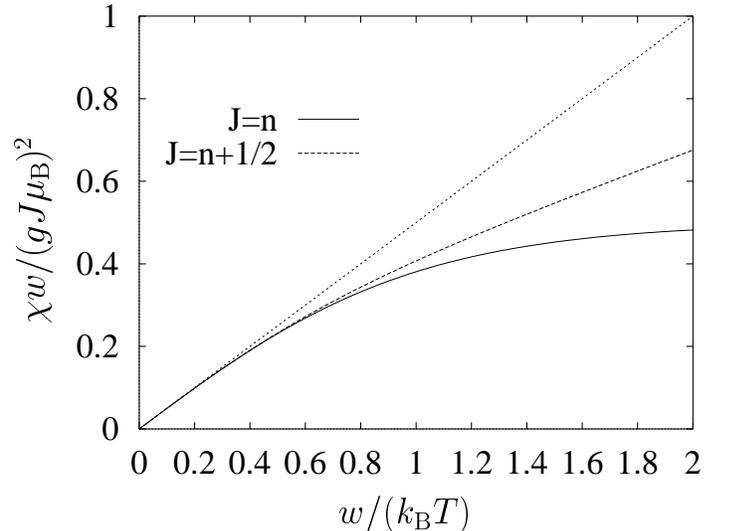}}
  \caption{Magnetic susceptibility versus inverse temperature. 
${\cal C}({\bf D}_4,2)$ configuration. 
The dotted line is the high temperature asymptote.}
  \label{fig:12}
\end{figure}

In the case of the ${\cal C}({\bf D}_N,2)$ configurations, $N=4,\,6$, 
the integer spin classes are non-magnetic and the half-integer ones are 
magnetic. 
The low temperature magnetic susceptibilities of these configurations 
can be found in the last column of Tables~\ref{tab:7},~\ref{tab:8} for the 
${\cal C}({\bf D}_4,2)$ and ${\cal C}({\bf D}_6,2)$ 
configurations respectively. 
A detailed temperature dependence of the magnetic susceptibility is 
shown on Fig.~\ref{fig:12} for the ${\cal C}({\bf D}_4,2)$ configuration. 

\begin{figure}
  \centerline{
\epsffile{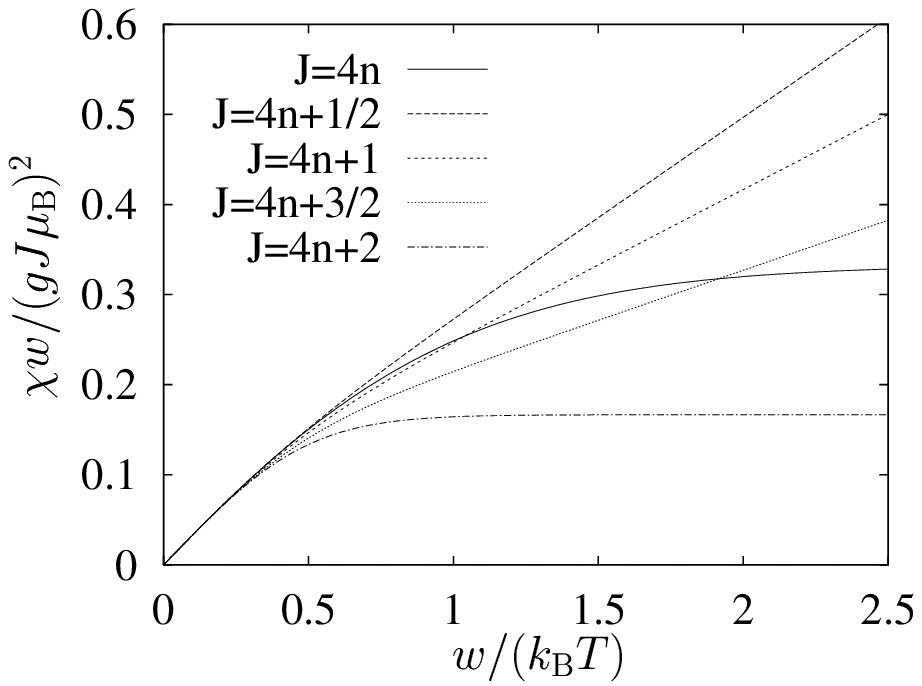}}
  \caption{Magnetic susceptibility versus inverse temperature. 
${\cal C}({\bf O},4)$ configuration.}
  \label{fig:13}
\end{figure}

\begin{figure}
  \centerline{
\epsffile{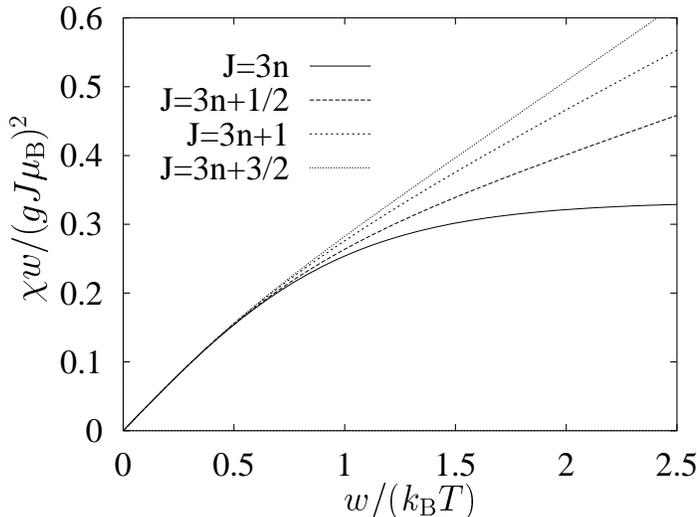}}
  \caption{Magnetic susceptibility versus inverse temperature. 
${\cal C}({\bf O},3)$ configuration.}
  \label{fig:14}
\end{figure}

The division into the classes of equivalence is more subtle for the 
high-order symmetry groups. Tables~\ref{tab:9},~\ref{tab:10},~\ref{tab:11}, 
and~\ref{tab:13} collect the low temperature susceptibilities for the 
${\cal C}({\bf O},4)$, ${\cal C}({\bf O},3)$, ${\cal C}({\bf Y},5)$, and 
${\cal C}({\bf Y},3)$ configurations respectively. 
Figs.~\ref{fig:13} and~\ref{fig:14} show 
the details of the transition 
from the Curie high temperature regime to the low temperature one for 
the ${\cal C}({\bf O},4)$ and ${\cal C}({\bf O},3)$ 
configurations respectively. 

For magnetic measurements it is important that the system is in thermal 
equilibrium and the range of temperature $Tk_{\rm B}<w$ is accessible. 
This requirement means that $w\sim \varepsilon_0 e^{-cJ}$ is not too small. 
On the other hand, $J$ must be not less than $\sim N/2$ to guarantee 
the validity of quasi-classical approximation. 
At a fixed lower limit for experimentally accessible temperature 
$T_l$ inequality $N<J<\ln(\varepsilon_0/(k_{\rm B}T))/c$ must be satisfied. 
For rare-earth ions $\varepsilon_0$ is the atomic scale of energy 
and $N=6,\,8$. 
It gives $T_l<\varepsilon_0e^{-4}\sim100$K which is easily satisfied. 
For La$_{1-x}$Ho$_x$Ni$_2$B$_2$C the estimated numerically $w$ is 
about 2K~\cite{Cho}, preliminary experimental results by 
D.~Naugle and coworkers give $w\approx1$K~\cite{Don}. 

Gd$^{+3}$ ion has zero orbital momentum, its anisotropy is caused by 
the relativistic spin-other-orbit interaction and corresponding 
$\varepsilon_0$ is about $10^{-4}$ time less than the atomic scale 
($\varepsilon_0\sim1\div10$K). The total spin of Gd$^{+3}$ ion 
$S=7/2$, not too large, but may be enough in the case of the 
tetragonal symmetry. The estimated value $T_l$ is between 0.1 and 1K. 

The anisotropy of a ferromagnetic cluster is induced mainly by its 
boundaries. The anisotropy energy has the same magnitude $\sim 1\div10$K 
per a site near the boundary. For the cluster as a whole this value 
must be multiplied by the number of atoms on one of the faces 
of the cluster (${\cal F}$) which depends on the cluster geometry. 
An estimate can be attained for the series of magic atom-number 
clusters~\cite{Jongh}: M$_{\cal N}$, ${\cal N}=13,\,55,\,147,\ldots$. 
These clusters are obtained by surrounding a core atom progressively 
with additional shells of atoms: $S_k=10k^2+2$, $k=1,\,2,\,3,\ldots$. 
This procedure can be done for icosahedral, decahedral, 
and cuboctahedral packings, 
which have 20, 15, and, 12 faces respectively.  
For ${\cal N}=55$ we find ${\cal F}\approx 42/16$ and 
$\varepsilon_0\approx3\div26$K. 
On the other hand $J\propto \zeta{\cal N}$. In the Gd cluster 
$\zeta\sim 0.5$ and for ${\cal N}=55$ we find $J\sim 27$. 
It is sufficiently large. The value of $w\sim\varepsilon_0e^{-cJ}$ 
with $c\approx 0.3$ is between 0.001 and 0.01K. 
For ${\cal N}=13$, $w$ ranges between 0.1 and 1K. 

\subsection{Spectral analysis.} 
\label{spectral_analysis}

The most straightforward experimental approach is the spectral analysis. 
The main difficulty on this way is that the scale of the splitting is very 
different for different systems and values of $J$. Nevertheless, we can 
expect that the spectral frequencies are either in the sub-millimeter 
or in the UHF range. Apart from the direct attenuation measurements, 
it is possible to apply EPR technique. It measures the splitting in 
magnetic field, {\em i.e.}, magnetic moment in some state. The advantage 
of this method is that it does not require too low temperatures. 
Certainly, its sensitivity drops with the growth of temperature, 
but not too fast.

\subsection{Oscillations of magnetization.}
\label{oscillations}

Let us consider many identical large spins placed into 
external magnetic field along one of the easy directions ($k$), 
sufficiently large to polarize them almost to saturation. 
If the field is switched off abruptly, each spin remains in the 
same state $|k\rangle$. Since $|k\rangle$ is not a stationary 
state, it will vary in time according to the Schr\"odinger picture: 
\be
|k,t\rangle = \sum_{j\alpha}|j\alpha\rangle\langle j\alpha|k\rangle
e^{-iE_jt/\hbar}.
\label{Shcrodinger}
\ee
Here $j$ labels sublevels of one $N$-plet and $\alpha$ labels states of 
the $j$th sublevel. 
It leads to oscillation of the magnetic moment 
along the $k$ direction in time: 
\bea
\label{moment}
M(t)&=&g\mu_{\rm B}J\sum_{k'}\cos\gamma_{kk'} \\
&\times& 
\sum_{j\alpha,j'\alpha'}
\langle j\alpha|k\rangle \langle j'\alpha'|k\rangle^{*} 
\langle j\alpha|k'\rangle^{*} \langle j'\alpha'|k'\rangle
e^{-i\omega_{jj'}t},
\nonumber
\eea
where $\gamma_{kk'}$ is the angle between the directions of classical 
angular momentum in the extrema $k$ and $k'$, and 
$\omega_{jj'}=(E_j-E_{j'})/\hbar$ is the transition frequency. 
All spins had the same initial state $|k\rangle$ at the moment when 
the field was switched off, therefore, their magnetic moment will rotate 
coherently creating the macroscopic rotating magnetization. Obviously, 
the rotation energy will dissipate. Let us estimate the attenuation 
time $\tau$. We assume that the spins are embedded into an insulator. 
Then only phonons leads to dissipation. The spin-phonon interaction 
energy can be written as follows:
\be
{\cal H}_{\mbox{s-ph}}=\lambda u_{\alpha\beta}J_\alpha J_\beta, 
\label{s-ph}
\ee
where $u_{\alpha\beta}$ is the deformation tensor. The value of the 
coupling constant can be estimated as $\lambda\sim\Delta/J^2$, where 
$\Delta$ is the energy difference of two oscillatory levels localized 
near one minimum of the potential $f({\bf J})$. A routine calculation 
leads to an estimate of the oscillations life-time $\tau$:
\be
\tau\sim\frac{\hbar\rho s^5}{\Delta^2\omega^3},
\label{relax}
\ee
where $\rho$ is the mass density of the matrix, $s$ is the sound velocity. 
For typical values $\rho=10$g cm$^{-3}$, $\Delta=10$K, 
$\omega=w/\hbar=10^{10}$s$^{-1}$, and 
$s=10^5$cm s$^{-1}$, we find $\tau\sim10^{-1}$s. The magnetic field must 
be switched off for a shorter time interval. It seems feasible. 
For Gd we estimated both $\Delta$ and $w$ by a factor of 10 smaller than 
the values we used for the above estimate. It gives the attenuation 
time $\tau$ in the range of few hours. 

In our estimate we assumed that the temperature $T$ is less or of 
the order of $w$. If it is much larger, the value of $\tau$~(\ref{relax}) 
must be multiplied by a small factor $\hbar\omega/(k_{\rm B}T)$. 
At a temperature 1K with $\hbar\omega\sim 0.01$ it changes $\tau$ from 
few hours to a minute, but still leaves this time long. 
Thus, the requirements for temperature is not too restrictive. 

Nevertheless, the observation of the macroscopic oscillations of 
magnetization may be obstructed because of inhomogeneous line broadening 
caused by the dipolar interaction~\cite{Slava}. Indeed, the random 
shift of the frequency due to the dipolar interaction is of the order: 
\be
\delta\omega\sim\frac{g^2\mu_{\rm B}^2J^2}{\hbar R^3}=
\frac{g^2\mu_{\rm B}^2J^2nx}{\hbar},
\label{shift}
\ee
where $R$ is the average distance between large spins, $x$ is their 
concentration per site, $n$ is the density of the matrix sites. 
For $g=2$, $J=3.5$, and $n=10^{22}$cm$^{-3}$ we find 
$\delta\omega\approx1.8\cdot10^{10}x\,$s$^{-1}$. 
For $x=0.001$ it three orders of magnitude less than 
$\omega\sim10^{10}$s$^{-1}$, but it destroys the coherence for the time 
interval $2\pi(\delta\omega)^{-1}\sim 10^{-7}$s. 
What can be observed after this interval of time 
is the noise in a rather narrow spectral range $\delta\omega$ 
given by Eqn.~(\ref{shift})
near the frequency $\omega$. The noise attenuates during the interval 
$\tau$ [Eqn.ps.~(\ref{relax})] 
after the pulse of magnetic field. Repeating the pulse of 
magnetic field periodically with the period $t<\tau$, one can 
maintain a permanent average level of the noise. 
Also, one can use this narrow-line noise to generate a coherent 
oscillations in a resonator. 
\section{Conclusion}
\label{section:Conclusion}

We have shown that large spins (total orbital momenta) $J$ 
placed into external fields of high symmetry group $G$ 
display unusual behavior of low-lying and high-lying
parts of spectra and magnetic susceptibility. 
These parts of spectra are represented by multiplets containing
$N(G,p)$ states each, where $N(G,p)$ is the doubled number of $p$-fold axes. 
Each multiplet is splitted into sublevels with multiplicities chosen from
dimensionalities of the irreducible representations of the point group $G$ and
determined by $G$, $J$ and $p$. 
The distances between sublevels in the multiplet
are proportional to $\exp{(-cJ)}$, whereas the distances between multiplets are
proportional to $1/J$. 
The multiplicities at a fixed $G$ and $p$ are periodic
functions of $J$ with the period $p$.
The relative distances between levels are also periodic functions on $J$, 
but their period is equal to a half of the number of the equivalent 
plaquettes formed by the tunneling trajectories and covering the unit sphere. 
Interesting exclusions are the configurations of the octahedron and 
icosahedron groups with $p=2$. In this cases 
the mutual arrangement of the levels is stochastic, though the
multiplicities remain fully deterministic. 

In all considered situations with exception of the tetrahedral and hexagonal
symmetry with in-plane easy directions, the change of large spin $J$ by 1 leads
to a drastic change in the spectrum and thermodynamic properties. We
demonstrated that at such a change the magnetic susceptibility can either
change its behavior from Curie law to saturation or change the coefficient in
Curie law.

Rather special phenomena appear near hyper-surfaces in the space of the
Hamiltonians which separate regions with different configurations of the 
extrema of the potential, {\em i.e.}, regions with different $N=N_1,\,N_2$. 
Directly on these
hypersurfaces the number of equivalent extrema is equal to $N_1+N_2$. 
Thus, in a narrow vicinity of the hyper-surface there appears 
a new "class of universality",
new set of sublevels with new multiplicities. Moreover, we expect a kind of
"turbulent" behavior of levels near these hyper-surfaces.

Given the classical Hamiltonian ${\cal H}({\bf J})$, one can indicate a value
$J_c({\cal H})$, starting from which the multiplicities are correctly 
determined by our theory. 
Though, this value is model-dependent, our numerical calculations show 
that $J_c\approx N(G,p)$.

All conclusions of the theory were checked numerically for a model Hamiltonian
of the cubic symmetry containing two invariants (two free parameters) up to 
$J=60$. The agreement for the relative distances between the levels 
is very good starting from $J\approx 20$. Multiplicities 
are well determined by our theory starting from $J\approx 12$ for the 
6- and 8-fold configurations and from $J\approx 16$ for the 
12-fold configuration. 

We proposed three classes of experimental systems 
which can display the predicted effects. 
One of them is represented by alloys with participation of two
lantanides or actinides, $R$ and $R'$, so that $R$ has zero orbital momentum 
and its
concentration is close to 1, whereas the element $R'$ has large $J$ and its
concentration is very small. In this way the configuration of large spin in a
symmetric environment is realized. Typical representatives are 
La$_{1-x}$Ho$_x$Ni$_2$B$_2$C (tetragonal environment) or 
Lu$_{1-x}$Dy$_x$Sb (cubic environment). 

The second class of systems are
metallic or metallo-organic clusters made from ferromagnetic metals. 
For such clusters symmetry can be not only octahedral, 
but also icosahedral, as it is for the cluster Fe$_{13}$. 
The clusters may have larger total spin than lanthanide and actinide atoms. 
In both cases we propose to measure spectrum of low-lying 
states (EPR or NMR measurements) and also to measure magnetization 
and magnetic susceptibility at low temperatures (about1-2K). 
Though experimental difficulties may arise on the way to 
realization of these experiments, we believe that the expected physical 
phenomena are worthwhile to study.   

The third class is magnetic dots used in experiments on magnetic 
tunneling~\cite{Barbara,Schuller}. 

Experimenters should choose optimal values of $J$ to ensure the validity 
of quasiclassical approach: reliable separation of the $N$-fold 
multiplets and simultaneously not too small values of the tunneling 
exponent $\exp(-cJ)$ with $c\leq 0.55$ for the cubic symmetry 
and $c\leq 0.29$ for the icosahedral symmetry. 

An interesting experimental and may be technical application of our 
system is the excitation of magnetic oscillations in a narrow spectral 
region by pulses of external magnetic field. The frequency of these 
oscillations range from $10^7$ to $10^{11}$Hz. 

\acknowledgments

We are thankful to E.~M\"uller--Hartmann and G.~S.~Uhrig who 
initiated this subject, and to  V.~V.~Dobrovitski 
for useful discussion. V.K. is grateful to I.~V.~Lavrinenko 
for the discussions on group theory. 
This work was supported by the NSF under 
Grant No. DMR-97-05182 and by the DOE under the Grant No. 
DE-F03-96ER45598. 
\appendix
\section{Characters of the main representation}
\label{appendix:characters}

The character of the identity transformation $E$ 
is trivial $\chi(E)=\dim(W)=N$.  
For a half-integer $J$, one has to find projective 
representations of the corresponding group with the factor set 
of $\{\pm1\}$ or, equivalently, linear representations of the 
group extended by a group $\{E,\,Q\equiv\exp(i2\pi J)E\}$  
(the so-called two-valued representations), where  
$Q$ is a rotation through an angle of $2\pi$ 
(here we adopt the notations of Ref.~\cite{LL-QM}). 
Obviously $\chi(Q)=\cos(2\pi J)\dim(W)=\cos(2\pi J)N$. 
Other elements having non-zero characters are the rotations with respect 
to the axes passing through the directions belonging to the set 
${\cal C}(G,p)$. 
Actually, it is sufficient 
to consider only one element from each class of conjugate elements. 
To calculate the corresponding characters we 
employ a following trick. 

\begin{figure}
  \centerline{
\epsfxsize=150pt \epsfysize=150pt 
\epsffile{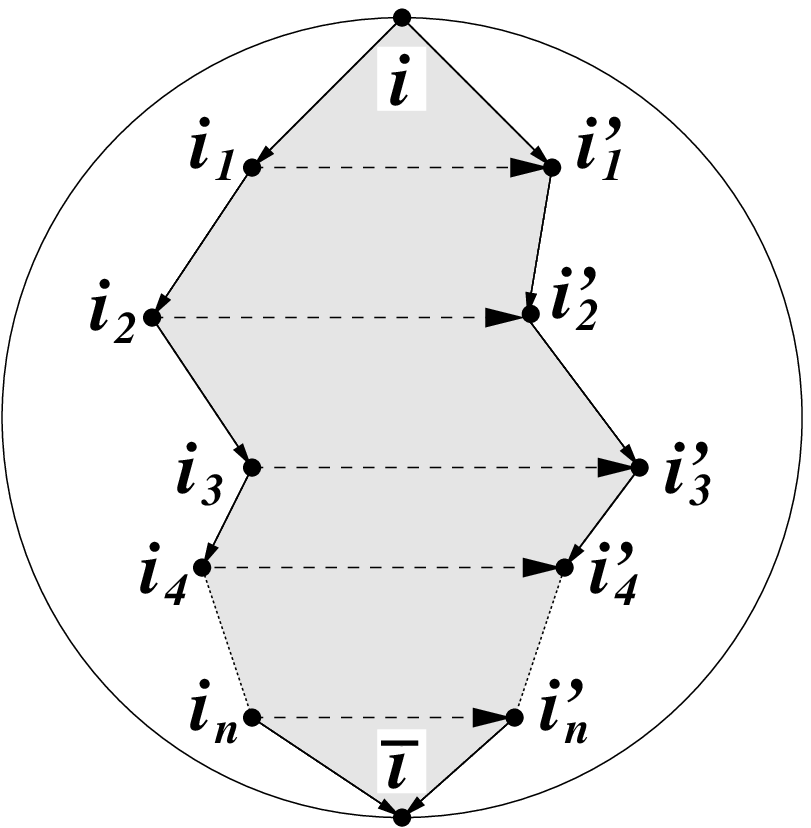}}
  \caption{Transformation of a tunneling path 
$i\rightarrow i_1\rightarrow\cdots\rightarrow i_n\rightarrow \bar\imath$ 
onto $i\rightarrow i_1^{\prime}\rightarrow\cdots\rightarrow 
i_n^{\prime}\rightarrow \bar\imath$. 
A solid angle of the filled area is $4\pi q/p$.}
  \label{fig:sec3_1}
\end{figure}

Let us consider an element of $P$ which, in our basis, 
corresponds to a rotation $C_p^q$ with respect to a $p$-fold axis. 
A set of rotations with respect to this axis forms a cyclic subgroup 
of $P$, and $q$ is a power of the generator of the subgroup $C_p$ 
(minimal non-trivial rotation); $q$ may take any integer value.
For a given configuration ${\cal C}(G,p)$ a non-zero character 
may occur only if $C_p^q$ leaves at least two of the extrema 
$i$ and $\bar\imath$ unmoved. It means that either 
the rotation axis passes through $i$ and $\bar\imath$, and $q\not=pn$ 
or the rotation is trivial: $q=pn$, $n$ is an integer.
Let us choose a tunneling path connecting $i$ and $\bar\imath$, 
and passing through intermediate nearest-neighbor extrema 
$i\rightarrow i_1\rightarrow\cdots\rightarrow i_n\rightarrow \bar\imath$ 
(see Fig.~\ref{fig:sec3_1}). 
Note, that some of the minima may coincide, that is, 
$i_j=i_k$ for some pairs $j\not=k$. 
The rotation $C_p^q$ transfers each extremum $i_j$ ($j=1,\,2,\ldots,n$) 
into $i_j^{\prime}$ leaving $i$ and $\bar\imath$ unchanged. 
The two paths form a closed loop on the sphere 
which subtends the solid angle of $4\pi q/p$. This fact leads to  
a relation for the oriented sum of the phases along the circuit:
\bea
\label{loop-sum}
&&\sum_{j=1}^{n-1}\left(\phi_{i_j,i_{j+1}}-
\phi_{i_j^{\prime},i_{j+1}^{\prime}}\right) \\
&+&\phi_{i,i_1}-\phi_{i,i_1^{\prime}}+
\phi_{i_n,\bar\imath}-\phi_{i_n^{\prime},\bar\imath}= 
J\frac{4\pi q}{p}.
\nonumber 
\eea
 
The same rotation $C_p^q$ transforms the phases in a following way: 
$\phi_{i_j,i_{j+1}}\mapsto\phi_{i_j^{\prime},i_{j+1}^{\prime}}$, 
$j=1,\,2,\ldots,n-1$, $\phi_{i,i_1}\mapsto\phi_{i,i_1^{\prime}}$, 
$\phi_{i_n,\bar\imath}\mapsto\phi_{i_n^{\prime},\bar\imath}$. 
To keep the phases unchanged (modulo $2\pi$) and the Hamiltonian 
invariant one should apply a gauge transformation 
${\cal U}=\mbox{diag}\{\exp(i\alpha_1),\ldots,\exp(\alpha_N)\}$, 
then the mappings turn into equations:
\bea
\phi_{i_j,i_{j+1}}&=&\phi_{i_j^{\prime},i_{j+1}^{\prime}}+
\alpha_{i_j}-\alpha_{i_{j+1}} \nonumber \\
\phi_{i,i_1}&=&\phi_{i,i_1^{\prime}}+
\alpha_{i}-\alpha_{i_1} \\
\phi_{i_n,\bar\imath}&=&\phi_{i_n^{\prime},\bar\imath}+
\alpha_{i_n}-\alpha_{\bar\imath} \nonumber
\eea
Substituting these equations into~(\ref{loop-sum}) gives:
$$
\alpha_{i}-\alpha_{\bar\imath}=J\frac{4\pi q}{p}, 
$$
which leads to:
\be
\alpha_{i_0}=J\frac{2\pi q}{p}+\delta,\,\,
\alpha_{\bar\imath_0}=-J\frac{2\pi q}{p}+\delta. 
\label{phases1}
\ee
Multiplying ${\cal U}$ with $\exp(-i\delta)$ removes the common factor 
$\delta$ in~(\ref{phases1}). 
Thus, the transformation ${\cal W}=\exp(-i\delta){\cal U}C_p^q$ 
of the configuration space has only two diagonal elements 
$\{{\cal W}\}_{ii}=\{{\cal W}\}_{\bar\imath\bar\imath}^{*}
=\exp(i2\pi Jq/p)$.
The character of the element ${\cal W}$ is $2\cos(J2\pi q/p)$.
All other characters are zeros since the corresponding rotations 
do not leave any state $|k\rangle$  of the configuration invariant. 
The characters of the representation do not depend upon the gauge: 
$\mbox{Tr}({\cal W}')=\mbox{Tr}({\cal U}{\cal W}{\cal U}^{\dagger})=
\mbox{Tr}({\cal W})$. 
Thus, one can study the reduction of the representation $W$ 
in any specific gauge without loss of generality.

\section{An example of a Hamiltonian for ${\cal C}({\bf O},4)$ configuration 
and calculation of its spectra.}
\label{appendix:A1}

In this appendix we consider a detailed construction of 
the reduced Hamiltonian for ${\cal C}({\bf O},4)$ configuration.
First we define the connectivity matrix of the system determining  
decide which minima ought to be connected by tunneling paths. 
This is usually done by connecting the nearest-neighbor minima. 
In some cases, however, the geometric closeness on the sphere is 
not a good criterion for connecting minima. 
A ``sure-fire'' criterion is the path integral approach which determines 
the amplitude of a spin transition from one localized state to another by 
summation of the contributions of all trajectories connecting the minima. 
This technique gives the exact solution of the problem, but it is very 
complicated. We could use instead a semiclassical  
method of finding trajectories with minimal imaginary classical action, 
but the action is not known itself. However, 
the symmetry of the system is of great help and is used throughout the 
paper to assess the connectivity structure. 
An example, in which the geometrically next-nearest-neighbors 
must be incorporated into the connectivity matrix together with the 
nearest-neighbors is discussed in Section~\ref{spectrum-O}. 


The connectivity matrix of ${\cal C}({\bf O},4)$ configuration is 
quite simple since only the geometric nearest-neighbors should be 
connected. 
For the minima 
enumeration given on Fig.~\ref{fig:1} the connectivity matrix is:
\be
\left|
\begin{array}{ccc}
\hat{\bf 0} & \hat{\bf 1}  & \hat{\bf 1} \\
\hat{\bf 1} & \hat{\bf 0}  & \hat{\bf 1} \\
\hat{\bf 1} & \hat{\bf 1}  & \hat{\bf 0} \\
\end{array}
\right|, \;\;
\hat{\bf 0}=\left|
\begin{array}{cc}
0 & 0  \\
0 & 0  \\
\end{array}
\right|, \;\;
\hat{\bf 1}=\left|
\begin{array}{cc}
1 & 1  \\
1 & 1  \\
\end{array}
\right| \ .
\ee

This is, actually, the Hamiltonian for $J=0 \pmod{4}$, 
without a multiplier of $w$, since the phase structure is absent 
or of no importance for the respective cases. 
For all other $J$'s one should find the twelve phases 
of the tunneling amplitudes. 
Five of the phases may be set to zero due to the gauge freedom 
with the only constraint that Eq.~(\ref{plaquette}) must be satisfied.
In our sample case, we null the following ones: 
$\phi_{1,3},\,\phi_{1,4},\,\phi_{1,5},\,\phi_{1,6},\,\phi_{2,5}$. 
The rest of the phases is obtained from the seven independent plaquettes 
(each plaquette gives an equation of type of Eq.~(\ref{plaquette})); 
these are: $\phi_{2,3}=-\pi J,\,\phi_{2,4}=\pi J,\,\phi_{2,6}=2\pi J,\,
\phi_{3,5}=\pi J/2,\,\phi_{3,6}=-\pi J/2,\,\phi_{4,5}=-\pi J/2,\,
\phi_{4,6}=\pi J/2$. 
Thus, the defined Hamiltonian is:
\be
{\cal H}_4^{\bf O}=\left|
\begin{array}{cccccc}
0 & 0            & 1             & 1                 & 1 & 1 \\
0 & 0            & e^{-i\pi J}   & e^{i\pi J}        & 1  & e^{i2\pi J} \\
1 & e^{i\pi J}   & 0             & 0                 & e^{i\pi J/2} & e^{-i\pi J/2} \\
1 & e^{-i\pi J}  & 0             & 0 & e^{-i\pi J/2} & e^{i\pi J/2}  \\
1 & 1            & e^{-i\pi J/2} & e^{i\pi J/2}      & 0 & 0 \\
1 & e^{-i2\pi J} & e^{i\pi J/2}  & e^{-i\pi J/2}     & 0 & 0 \\
\end{array}
\right| \ .
\label{O6f}
\ee

Finally, one needs to find the eigenvalues of the Hamiltonian. 
The diagonalization can be performed by any symbolic solving system or 
even ``manually'' since the Hamiltonian can be factorized. 
Also, a trick of a purely geometric origin can be used: if one views 
Hamiltonian ${\cal H}$ 
($w=1$ and $|({\cal H})_{ij}|=1$ if $({\cal H})_{ij}\not=0$) 
as a weighted connectivity matrix of a graph, then:
\be
\sum_{i}g_iE_i^n=\mbox{Tr}{\cal H}^n=2\sum_{j}\sum_{k_j}\cos(\Omega_{k_j})
\equiv I_n,
\label{trick}
\ee
where $n=0,\,1,\,2,\ldots$, $E_i$ is the $i$th distinct eigenvalue of 
${\cal H}$ of multiplicity $g_i$, 
the first sum on right-hand side is over the vertices of the graph, 
the second is over all closed loops of $n$ walks running through vertex $j$ 
and $\Omega_{k_j}$ is the flux passing through the $k_j$th loop. 
If $({\cal H})_{ii}\not=0$, $i=1,\ldots,N$, 
some extra weights need to be applied for each loop. 
An advantage of formula~(\ref{trick}) is that $I_n$ are gauge invariant 
since all $\Omega_{k_j}$ are gauge invariant. 
Applied to Hamiltonian~(\ref{O6f}) for $n<4$, the formula yields:
\begin{eqnarray}
\label{trick3}
\sum_i g_i=6, && \sum_i g_iE_i=0, \\
\sum_i g_iE_i^2=24, && \sum_i g_iE_i^3=48\cos(\pi J/2).\nonumber 
\end{eqnarray}

Let us find the spectrum of Hamiltonian~(\ref{O6f}) for $J=0$.
We know from  Section~\ref{O} of Section~\ref{section:Group} 
that $g_1=1$, $g_2=2$, and $g_3=3$. 
Too, it is clear that $E_1=4$ (sum of the matrix elements in each 
row or column of Hamiltonian~(\ref{O6f}) is equal to 4 for $J=0$). 
Two equations for $E_2$ and $E_3$ are:
\be
4+2E_2+3E_3=0,\,\,16+2E_2^2+3E_3^2=24.
\label{Of6J0}
\ee
Solving~(\ref{Of6J0}) and checking the roots against the last equation 
of~(\ref{trick3}) gives $E_2=-2$ and $E_3=0$. 
For $J=2$, the spectrum is inverted (see Section~\ref{section:Spectrum}): 
$E_1=-4$, $E_2=0$, and $E_3=2$. For all other $J$'s the solution is 
trivial since there are only two distinct eigenvalues.

\section{Low temperature magnetic susceptibility.}
\label{appendix:B}

For high temperatures $k_{\rm B}T>w$, the moments are purely 
classical and show Curie magnetic susceptibility:
\be
\chi_{\rm C}=(\mu_{\rm B}gJ)^2/(dk_{\rm B}T).
\label{Curie}
\ee
where $d$ is the dimensionality of the system, 
$\mu_{\rm B}$ is the Bohr magneton, 
$k_{\rm B}$ is the Boltzmann constant, and $g$ is the gyro-magnetic ratio. 
For low temperatures $k_{\rm B}T<w$, the quantum effects change 
the response drastically. The susceptibility saturates 
for the classes without magnetic moment in the ground state to the value:
\be
\chi_s=\frac{1}{g_0}\sum_{i=1}^{g_0}\chi_{i},
\label{suscept_zero}
\ee
where $g_0$ is the degeneracy of the ground state and 
$\chi_{i}$ is the susceptibility of the $i$th member of the 
ground-state multiplet. 
For the classes with the magnetic moment in the ground state 
Curie susceptibility persists at $k_{\rm B}T<w$, but its slope 
is different:
\be
\chi_l=\frac{1}{g_0}\sum_{i=1}^{g_0}m_{i}^2/(k_{\rm B}T),
\label{suscept}
\ee
where $m_{i}$ is the moment of the $i$th member of the 
ground-state multiplet. 


\section{On the intersection of the multiplets of 
${\cal C}({\bf O},4)$ and ${\cal C}({\bf O},3)$.}
\label{appendix:O14}

In a close vicinity of a minimum of 
${\cal C}({\bf O},4)$ and ${\cal C}({\bf O},3)$ configurations, 
Hamiltonian~(\ref{hamilton2}) has the following form:
\bea
\label{hamilton2_4}
{\cal H}_2^{{\bf O}4}=-a-b+(2a+3b)(x^2+y^2), \\
\label{hamilton2_3}
{\cal H}_2^{{\bf O}3}=-\frac{1}{3}a-\frac{11}{9}b+\frac{4}{3}(4b-a)(x^2+y^2), 
\eea
where $x$ and $y$ are local Cartesian coordinates. 
Treating these terms as the effective potential energy of 
the quantum-mechanical problem, 
one can identify it with that for a 
2-dimensional isotropic harmonic oscillator 
(the kinetic energy is due to the Wess-Zumino term~\cite{WZ} 
or Berry phase, its exact form is of no importance here; it suffices to know 
that this term is identical for both~(\ref{hamilton2_4}) 
and~(\ref{hamilton2_3}), 
thus, providing identical effective masses $M$). 
The potential energies of the two harmonic oscillator problems 
are equal on a line $b=3a$. The squares of the effective frequencies 
are $22a/M$ and $88a/(3M)$ ($a>0$), respectively 
for ${\cal C}({\bf O},4)$ and ${\cal C}({\bf O},3)$, on this line. 
Hence, the energy of the ground 
as well as the spacing between successive levels is larger for 
${\cal C}({\bf O},3)$ configuration on this line. 
These arguments qualitatively explain the deviation 
of the line $b=3a$ from the point of the level crossing towards the 
6-fold coordination region. 

The boundary of the transition from one configuration to another. 
(a line in our 2-dimensional $a$-$b$ space) marks a singularity in  
the ``flow'' of the level multiplicities across the parameter space. 
The levels of two different coordinations must match exactly at this surface. 
We find the spectra of this intermediate configuration 
in the case of $N=N({\bf O},4)+N({\bf O},3)=14$. 

The surface of the sphere is covered with twelve congruent even-sided 
plaquettes. Hence, the period of spectra is $J=6$ and all spectra 
are symmetric (see Section~\ref{section:Spectrum}). 
The spectra of the 14-fold configuration are collected in Table.~\ref{tab:d1} 
for the non-equivalent $J$'s. 

\begin{table}
\caption{The spectra of a ``hybrid'' 
${\cal C}({\bf O},4)+{\cal C}({\bf O},3)$ configuration. 
}
\begin{tabular}{lll}
$J$ & Eigenvalues(degeneracies) \\ \hline
0 & $\pm2\sqrt3w\,(1,1)$, $\pm2w\,(3,3)$, $0\,(6)$  \\
1 & $\pm(1+\sqrt3)w\,(3,3)$, $\pm(1-\sqrt3)w\,(3,3)$, $0\,(2)$  \\
2 & $\pm\sqrt6w\,(2,2)$, $\pm2w\,(3,3)$, $0\,(4)$            \\
3 & $\pm2w\,(6,6)$, $0\,(2)$  \\
1/2 & $\pm w\sqrt{6+2\sqrt3}\,(2,2)$, $\pm w\sqrt{3-\sqrt3}\,(4,4)$, $0\,(2)$ \\
3/2 & $\pm w\sqrt6\,(4,4)$, $0\,(6)$ \\
5/2 & $\pm w\sqrt{6-2\sqrt3}\,(2,2)$, $\pm w\sqrt{3+\sqrt3}\,(4,4)$, $0\,(2)$ \\
\end{tabular}
\label{tab:d1}
\end{table}


\end{document}